\documentclass[showpacs,amsmath,amstex,amssymb,mathfonts,pre]{revtex4-1}
\usepackage{graphicx}

\usepackage{tipa}
\usepackage{bm}

\usepackage{amsmath}
\usepackage{amssymb}
\usepackage{amsthm}
\usepackage{amsfonts}
\usepackage{float}
\usepackage{enumitem}
\usepackage{makecell}
\usepackage{multirow}
\usepackage{verbatim}

\begin{document}

\title{Dynamics of Taxi-like Logistics Systems: Theory and Microscopic Simulations}

\author{Bo Yang$^{1,2}$ and Qianxiao Li$^{2,3}$}
\affiliation{$^1$Division of Physics and Applied Physics, Nanyang Technological University, Singapore 637371.\\
$^2$ Computing Science Department, Institute of High Performance Computing, A*STAR, Singapore, 138632.\\
$^3$ Department of Mathematics, National University of Singapore, 117543.
}
\date{\today}
\begin{abstract}
We study the dynamics of a class of bi-agent logistics systems consisting of two types of agents interacting on an arbitrary complex network. By approximating the system with simple microscopic models and solving them analytically, we reveal some universal dynamical features, and propose applications of such features for system optimisations. Large scale agent-based numerical simulations are also carried out to explore more realistic and complicated systems, with interesting emergent behaviours that can be well understood from our analytical studies. Using the taxi system as a typical logistics system with commuters and empty taxis as two types of agents, we illustrate two dynamical phases with distinct behaviours, separated by a phase boundary that can be identified as the optimal number of taxis for a particular taxi system. These features and the tuning of the optimal number of taxis can be applied to various situations, including taxi systems allowing real-time dynamical ride-sharing.
\end{abstract}

\maketitle

%%%%%%%%%%%%%%%%%%%%%%%%%%%%%%%%%%%%%%%%%%%%%%%%%%%%%%%%%%%%%%%%%%%%%%%%%%%%%%%%
\section{INTRODUCTION}
Logistics management involving flow of goods from origins to destinations, via delivery agents moving in a regular or complex network, is an interesting example of complex systems in which a large number of components interact with each other in strongly non-linear ways\cite{laporte,bock}. While the rules of interaction are generally quite simple, the dynamics of such systems can be diverse and rather unpredictable, leading to universal emergent behaviours that can be exploited for system optimisation\cite{yang}. From a practical point of view, the management of such logistics system mainly involves efficient resource allocation in both spatial and temporal domains, especially when there is a need to respond to goods with origins and destinations generated on a real time and dynamical basis\cite{bilge,zhong,laporte2,jaillet,delage,chowjy,yej}.

As an example, the taxi system consists of mainly two types of agents - commuters and taxis - interacting over the domain of an urban road network. The main task for the taxis is to deliver commuters from their origins to their respective destinations, and the available resource is the fleet of empty taxis at any moment of the day. The management of such resources is given by the dynamical allocation of the empty taxis over the entire road network, subject to factors such as the connectivity of the road network, real time traffic conditions, and vehicle travelling speeds. In principle, such management can be effectively implemented by controlling the total number of taxis in the system, as well as the routing of individual taxis\cite{mohsen,reisman}. The routing from origins to destinations for occupied taxis is generally straightforward, normally with the shortest path taken subject to certain constraints (e.g. traffic conditions or tolls, etc.). Highly nontrivial algorithms can be developed if ride-sharing is involved, where occupied taxis may also be available to additional commuters based on judicious route-matching algorithms\cite{yang,koening,roorda,reijers,dchen}. On the other hand, routing for empty taxis is nontrivial even without ride-sharing, with several approaches in customer searching and demand predictions\cite{wong,damas,yanghai2,kendall,ramezani,chen,wong2,yux,zha}. The optimal routing algorithm also depends on the available communication technologies. While in olden days road-side hailing is the only option for empty taxis to meet commuters, nowadays advanced real-time taxi booking via smartphones can be easily implemented, so that empty taxis have additional information of where the commuters are waiting and where their destinations are, before selecting an efficient routing strategy.

There are also other similar logistics systems, consisting of agents interacting in analogous ways. Instead of commuters, picking up and delivering goods efficiently on a large scale is a common challenge for maritime port management, storage management and postal delivery, etc. It is becoming more prevalent with the popularity of online-to-offline services (e.g. grocery delivery) and the advances in automated technologies (e.g. autonomous vehicles and drones). A highly efficient and decentralised delivery network/system could be an important part of the overall logistics management in the near future, especially when the demand is generated in a real-time and less predictable manner. In general there can be many practical complications and constraints for such delivery systems and on-demand services\cite{ham,sundar,changsun}, and sophisticated pick-up and delivery algorithms could be developed especially in skip delivery problems (SDP) or split delivery vehicle routing problems (SDVRP)\cite{archetti,speranza}. In this work, however, we use abstract models to simplify these complications, with the aim of finding universal features resulting from the dynamics of such systems that are unaffected by certain details systematically ignored in our analysis and simulations. Understanding such universal features could be useful in benchmarking and optimising these logistics systems with various resource allocations, routing and delivery algorithms.

Throughout this paper we denote the agents that are delivered from origins to destinations as $\mathcal G$, or $\mathcal G$ agents; and the delivery agents that moves $\mathcal G$ agents around as $\mathcal L$, or $\mathcal L$ agents. While this paper does not focus on actual system optimisation strategies themselves, here we give a brief discussion on a number of optimisations the analysis in this paper can be useful for. The optimisation of the number of the $\mathcal L$ agents, as well as the routing algorithm, can depend on a number of factors. In general, we would like to minimise the number of $\mathcal L$ agents needed to save cost. In addition, a good routing algorithm should also minimise the average distance travelled by $\mathcal L$ agents, which can be physically related to the ``fuel cost" or the ``maintenance cost". For transportation systems, reducing the number of $\mathcal L$ agents and the number of trips can also help reduce traffic congestions. If rewards are given for every $\mathcal G$ agent delivered, we can also try to maximize the total or average earnings of $\mathcal L$ agents. From the perspective of $\mathcal G$ agents, the most important factor is the average waiting time; for every $\mathcal G$ agent generated in the system, we would like an $\mathcal L$ agent to pick it up as soon as possible. If pooling or ride-sharing is involved, the trip time for a $\mathcal G$ agent to be with an $\mathcal L$ agent not only depends on the intrinsic origin and destination for $\mathcal G$ agents, but also on possible detours for additional pick-up/drop-off of the shared parties. It is thus also sensible to minimise the total travel time (which is the sum of the waiting time and the trip time), or the total cost of the trip (i.e. for taking a taxi). The overall utility function of the logistics system can be one or a combination of a number of factors mentioned above.

In this paper, we do not focus on how to optimize such logistics systems from the perspectives mentioned in the previous paragraph, as this will be discussed in detail elsewhere\cite{yangunpublished}. Instead, we give a well-defined mathematical formulation of the dynamics of such complex systems, followed by employing both analytical and numerical tools to understand and formulate universal and important features of the dynamics of such systems. These features will be useful for benchmarking system efficiency, predicting qualitatively different behaviours of the agents, as well as guiding specific optimisations of these logistics systems. Given the non-linear interactions between agents over the complex network, the challenge is to solve such systems with limited analytical tools. Simplification of the mathematical formulation by ignoring some detailed agent behaviours, combined with macroscopic phenomenological reformulation of the problem where analytical treatment is possible, can lead to useful approximate solutions that captures the essential dynamics. It is, however, indispensable to have microscopic agent based simulations that are as realistic as possible, to evaluate and validate these approximate solutions. Such microscopic simulations are also very useful in exploring general behaviours of the system dynamics under various different scenarios.

This paper will be organised as follows: in Sec.~\ref{math} we formulate a network based logistics system with two types of agents in precise mathematical language; in Sec.~\ref{theory} we study mathematical models with simple settings but non-trivial behaviours, that capture essential features of the dynamics of the network-based logistics systems with two types of agents; in Sec.~\ref{numerics} we study numerically a particular example of the taxi systems, revealing various useful dynamical features that can be understood via the mathematical analysis in Sec.~\ref{math}; in Sec.~\ref{dphase} we focus on the two dynamical phases of the taxi systems, showing that even with the additional dynamics involving ride-sharing, the formulation of different phases from formal and numerical analysis of the bi-agent logistics system can be useful for predicting complicated behaviours affecting the quality of the taxi services; in Sec.~\ref{summary} we summarise our work and discuss about the outlooks.

\section{Mathematical Formulation}\label{math}

At a more abstract level, the logistics systems we focus on is a bi-agent system consisting of two types of agents, with non-trivial dynamics on a potentially complex network. We represent the network as a graph $G\left(N,E\right)$, where $N$ is the collection of nodes of the graph, and $E$ is the collection of directed, weighted edges connecting the nodes. The graph can also be fully represented by an adjacency matrix $\mathcal A$. For nodes $i,j$, $\mathcal A_{ij}$ is non-zero if there is an edge directed from $i$ to $j$. This is the case if and only if an agent can move from $i$ to $j$ without passing through any other nodes (so the $j^{\text{th}}$ node is directly connected from $i^{\text{th}}$ node). The matrix element $\mathcal A_{ij}$ gives the \emph{inverse} of the time it takes for an agent to move from $i$ to $j$. Following this definition, we take $\mathcal A_{ij}^{-1}$ to be a positive integer, in the unit of the smallest time resolution of the system (e.g. one second). Thus $\mathcal A_{ij}$ can also be time-dependent, and if different agents have different velocities, each agent will have its own corresponding adjacency matrix. $\mathcal A_{ij}=0$ implies no edge between the two nodes, as the time it takes to travel from $i$ to $j$ is infinity.  

We denote the two types of agents as $\mathcal G$ and $\mathcal L$; the corresponding rate of generation of the agents at $i^{\text {th}}$ node is given by $g_i$ and $l_i$ respectively, and both in principle can be time dependent. One or both agents can move from one node to another via edges, and when an agent $\mathcal G^m$ and an agent $\mathcal L^n$ (note the upper indices give the agent index, while the lower indices give the node index) meet at the same node, they ``annihilate" each other in pairs, and leave the system. Let $N_{\mathcal G}$ and $N_{\mathcal L}$ be the number of agents of $\mathcal G$ and $\mathcal L$ respectively, the most important dynamics of such logistics systems is the time dependence of these two quantities, denoted by $N_{\mathcal G}\left(t\right)$ and $N_{\mathcal L}\left(t\right)$.

The motion of the agents is described by an policy matrix $\mathcal P^n_{ij}$, where $n$ is the index of the agents. For an agent at node $i$, its probability of moving towards a neighbouring node $j$ at the next time step is given by $\mathcal P^n_{ij}$; thus we have $\sum_j\mathcal P^n_{ij}=1$, and $\mathcal P^n_{ij}\neq 0$ if and only if $\mathcal A_{ij}\neq 0$. In principle, each agent can have its own policy matrix. If the motion of the agents are controlled in a centralised manner, then the policy matrix can differ from one agent to another based on the control algorithm, and can also be time dependent. On the other hand, for decentralized systems in which agents make their own decisions, all agents of the same type ($\mathcal G$ or $\mathcal L$) can follow the same policy matrix. If that is the case, we omit the upper indices and simply denote the policy matrix as $\mathcal P_{ij}$. It is also common for agents of the same type to follow the same policy matrix under certain conditions, and individualised policy matrices under other conditions (e.g. the implementation of booking policies in taxi systems, as we will discuss in Sec.~\ref{numerics}).

The dynamics of this interacting system is thus completely defined by $\mathcal A_{ij}, g_i,l_i, \mathcal P_{ij}^n$, and the annihilation rule between the two types of agents. We would like to emphasize here that for realistic systems, $g_i$ and $l_i$ may not be independent. Moreover, the generation of both $\mathcal G$ and $\mathcal L$ agents may be correlated across graph nodes. As we shall see in Sec.~\ref{ps}, the ``annihilation" between agents from $\mathcal G$ and $\mathcal L$ can also correspond to the formation of ``bound states" with non-trivial dynamics, with the ``bound state" delivering the bound $\mathcal G$ agent in it from its origin to its destination. At the destination the $\mathcal G$ agent leaves the system, while the $\mathcal L$ agent in the ``bound state" re-enters the system. If the latter is the only physical way for agents from $\mathcal L$ to be generated in the system, then $l_i$ depends on $\mathcal A_{ij}, \mathcal P_{ij}$ and the origin/destination distribution of $\mathcal G$. Theoretically, it is useful to treat the dynamics of the ``bound states" as hidden, and $l_i$ as dynamical and tunable. The characteristic dynamical behaviours can be systematically studied both analytically and numerically, as we will show in Sec.~\ref{theory} and Sec.~\ref{numerics}.

\subsection{Taxi System as a Special Case}\label{ps}

In a taxi system, the network $\mathcal A_{ij}$ corresponds to the road network, and the two types of agents are the \emph{empty} taxis ($\mathcal L$) and the commuters ($\mathcal G$). The nodes are locations where commuters can board or alight the taxis, and the ``annihilation" process corresponds to a commuter being picked up by an empty taxi when they meet at the same node. The commuters are stochastically generated at different nodes in the road network, with probability $g_i$ per time step either obtained from historical empirical data, or artificially synthesized. When a commuter is picked up by an empty taxi, both agents will disappear from the system. Empty taxis will re-emerge at different nodes according to $l_i$, corresponding to the roaming of the empty taxis and the spatio-temporal distribution of the destinations of the commuters, where commuters alight (but not re-enter the system) and the taxis become empty again.

In this particular system, we treat commuters as non-mobile: once they are generated, they stay at the nodes with a trivial policy matrix $\mathcal P_{ij}=\delta_{ij}$. The empty taxis, on the other hand, are mobile with a non-trivial policy matrix. The efficiency of the taxi system depends strongly on how good this policy matrix is, corresponding to the strategy of empty taxis in anticipating where potential commuters will be in the road network.

The taxi system is also a typical example with non-trivial ``hidden dynamics" of ``bound states" representing an occupied taxis with a boarded commuter. In general, an occupied taxi will go along the shortest path from the origin to the destination of the passenger. Thus, part of $l_i$ depends on factors including $g_i$ and the distribution of the destinations, which we denote as $M_{ij}$. Physically, $M_{ij}$ is the probability that a commuter boarding at node $i$ will alight at node $j$, so we have $\sum_jM_{ij}=1$. In our theoretical treatment in Sec.~\ref{theory}, we ignore the dynamics of the occupied taxis, and treat the emergent $l_i$ as a phenomenological input. For microscopic agent based simulations in Sec.~\ref{numerics}, on the other hand, the dynamics of both empty and occupied taxis are fully accounted for, and meaningful comparisons between theory and simulations can be made.

We also would like to comment here that the dynamics of the occupied taxis is generally quite straightforward if no ride-sharing is involved, as is the case in the most part of this paper. Real-time adaptive ride-sharing is a very important area of research, where the routing of occupied taxis and the route matching of different commuters can be highly non-trivial (see ref.\cite{yang} and the references therein). We will analyze this more complex system in details elsewhere; a number of theoretical and numerical results in this work will also lay the foundation to a systematic characterization of taxis with dynamical ride-sharing, as we will illustrate in Sec.~\ref{dphase}.
\begin{table}[htb]
\centering
\begin{tabular}{| c | c |}
\Xhline{3\arrayrulewidth}
&\\
Notations &  Physical meaning\\
&\\
\hline
&\\
$g_i$ & Rate of generation of $\mathcal G$ agent at node $i$\\
&\\
\hline
&\\
$l_i$& Rate of generation of $\mathcal L$ agent at node $i$\\
&\\
\hline
&\\
&The OD matrix, the matrix element gives the probability\\
$M_{ij}$& for the commuter generated at node i to have\\
&destination at node $j$\\
&\\
\hline
&\\
$\mathcal A_{ij}$& Weighted, directed adjacency matrix for the network,\\
& the matrix element gives the inverse of travel time across the edge\\
&\\
\hline
&\\
& Policy matrix, the matrix element gives the probability\\
$\mathcal P^n_{ij}$& for the agent at node $i$ to move to node $j$.\\
& The subscript is the agent index, which can be omitted\\
& if all agents behave the same way.\\
&\\
\hline
\end{tabular}
\caption{Components and notations for the two-species logistics system.}
\label{t}
\end{table}

\section{Theoretical Modelling and Analysis}\label{theory}

Given the adjacency matrix of the network $\mathcal A_{ij}$, the spatiotemporal distribution of the rate of generation of the two agents given by $g_i, l_i$, and the policy matrix $\mathcal P_{ij}$ governing the movement of the agents, we would like to analyse both the equilibrium and non-equilibrium dynamics of the agents in the system. In particular, one can calculate many useful quantities (e.g. average waiting time, average travel distances, etc.) from $N_{\mathcal G}$ and $N_{\mathcal L}$, the number of the two types of agents as a function of time.

It is useful to first look at the simple model of a single node with fixed probabilities $g$ and $l$, where for each time step $g$ is the probability of generating a $\mathcal G$ agent, while $l$ is the probability of generating an $\mathcal L$ agent. The latter is equivalent to removing a $\mathcal G$ agent at the node if and only if there is one present, since a $\mathcal G$ agent and an $\mathcal L$ agent annihilate each other. We also assume that agents of type $\mathcal G$ can accumulate at the node, while agents of type $\mathcal L$ do not; thus if an $\mathcal L$ agent is generated at the node with no $\mathcal G$ agents present, this $\mathcal L$ agent will be removed at the end of the time step. This corresponds to the case that in the full network, $\mathcal G$ agents do not move, while $\mathcal L$ agents roams in the network with some policy matrix.

One should also note that given our system is discrete in the time domain, both $g$ and $l$ (as well as $g_i$ and $l_i$ in the rest of this paper) are  the average number of agents of $\mathcal G$ and $\mathcal L$ generated in $\Delta t$, where $\Delta t$ is the time interval between consecutive time steps. Thus $g$ and $l$ scale with $\Delta t$, and there is thus no constraint that $g$ and $l$ cannot be larger than $1$. For $g>1$ or $l>1$, implementations in numerical simulations can also be carried out straightforwardly by rescaling $\Delta t$, while keeping the physically relevant probability densities $g/\Delta t, l/\Delta t$ constant. Alternatively, for small systems one can employ the Gillespie algorithm\cite{gillespie1977exact} to simulate the continuous-time Markov chain directly. For convenience, the notation used in this section is summarized in Table.~\ref{tt}.

\begin{table}[htb]
        \centering
        \begin{tabular}{| c | c |}
        \Xhline{3\arrayrulewidth}
        &\\
        Notations &  Physical meaning\\
        &\\
        \hline
        &\\
        $g$ & Rate of generation of $\mathcal G$ agent at the node\\
        &\\
        \hline
        &\\
        $l$& Rate of generation of $\mathcal L$ agent at the node\\
        &\\
        \hline
        &\\
        $\Delta t$& Time interval between consecutive steps in simulation\\
        &\\
        \hline
        &\\
        $N_{\mathcal{G}}(t)$& Number of $\mathcal{G}$ agents at time $t$ at the node\\
        &\\
        \hline
        &\\
        $N_{\mathcal{L}}(t)$& Number of $\mathcal{L}$ agents at time $t$ at the node\\
        &\\
        \hline
        &\\
        $\bar{\omega}_T$& Expected average weighting time of $\mathcal{G}$ agents up to time $T$\\
        &\\
        \hline
        \end{tabular}
        \caption{Additional notations for the theoretical analysis in Section \ref{theory}.}
        \label{tt}
\end{table}

\subsection{The Single Node Model}\label{snm}

We now look at the simple model with only one node, with the associated $l$ and $g$ both satisfying $0\leq l+g\leq 1$, which can always be achieved by rescaling $\Delta t$. Similar problems with the constraint that $l>g$ has been studied before\cite{bailey}. Our analysis here is more general, and we can view the dynamics of the accumulated waiting time as a Markov chain with state space $\mathbb{N}=\{0,1,2,\dots\}$. More concretely, we have the number of $\mathcal G$ agents at the node at time t, $N_{\mathcal G}\left(t\right)\in\mathbb{N}$. We thus have the initial condition $N_{\mathcal G}\left(0\right)=0$ and the recursive relation as follows:
\small
\begin{align}
        \begin{split}
                &p(t+1,n)  \\
                =&
                \begin{cases}
                        g p(t,n-1) + l p(t,n+1) + (1-g-l) p(t,n) & n\geq 1\\
                        l p(t,n+1) + (1-g) p(t,n) & n=0,
                \end{cases}
        \end{split}
\end{align}
\normalsize
where we have defined $p(t,n):=\mathbb{P}[N_{\mathcal G}\left(t\right)=n]$, the probability of having $n$ $\mathcal G$ agents waiting at the node at time step $t$. Of course, we have the initial condition $p(n,t)=\delta_{n,0}$. By computing the eigenvalues and eigenvectors associated with the transition probability matrix, we can derive the following explicit form of $p(t,n)$:
\begin{align}
        p(t,n) &=
        \begin{cases}
            \frac{l-g}{l}{\left( \frac{g}{l} \right)}^{n}
            + {\left(\frac{g}{l}\right)}^{\frac{n}{2}} I_n(g,l,t) & g<l \\
            {\left(\frac{g}{l}\right)}^{\frac{n}{2}} I_n(g,l,t) & g\geq l
        \end{cases}
        \label{eq:p_soln}
\end{align}
where
\begin{align}
        I_n(g,l,t):= \frac{2g}{\pi}
        \int_{0}^{\pi}  \frac{{\lambda(g,l,\theta)}^t}{1 - \lambda(g,l,\theta)}
        f_n(g,l,\theta) \sin\theta d\theta\label{laplace}
\end{align}
and
\begin{align}
        \begin{split}
                \lambda(g,l,\theta) &:= 1-g-l+2\sqrt{gl}\cos\theta \\
                f_n(g,l,\theta) &:= \sin[(n+1)\theta] - \sqrt{\tfrac{l}{g}} \sin(n\theta)
        \end{split}
\end{align}
Furthermore, by performing asymptotic analysis via Laplace's method on the integral in Eq.(\ref{laplace}) as {$t\rightarrow\infty$, we find that
\begin{align}
        \mathbb{E}[N_{\mathcal G}\left(t\right)] =
        \begin{cases}
            (g-l)t + o(t) & g > l \\
            \mathcal{O}(t^{\frac{1}{2}}) & g = l\label{asymptotics} \\
            \mathcal{O}(1) & g < l
        \end{cases}
\end{align}
where $\mathbb{E}[N_{\mathcal G}\left(t\right)]=\sum_{n=0}^\infty np\left(t,n\right)$ is the statistical average of the number of waiting $\mathcal G$ agents at the node at time $t$. This signals a second-order phase transition at the line $g=l$.

An important quantity in this system is the average waiting time of the $\mathcal G$ agents, which is averaged over the total number of $\mathcal G$ agents being generated over the simulation time $T$. Its formal expression is given by the following, with the associated asymptotics derived from Eq.(\ref{asymptotics}), as:
\begin{align}
       \bar\omega_T=\frac{1}{Tg}\sum_{t=1}^{T}\mathbb{E}[N_{\mathcal G}\left(t\right)]=
        \begin{cases}
            \mathcal O\left(T\right) & g > l \\
            \mathcal{O}(T^{\frac{1}{2}}) & g = l \\
            \mathcal{O}(1) & g < l
        \end{cases}
\end{align}
To carry out non-asymptotic analysis, we write down the explicit expression for $\bar\omega_T$
\begin{align}
        \bar\omega_T
        =\frac{1}{Tg}\sum_{t=1}^{T} \sum_{n=0}^{\infty} n p(t,n)\label{singlenode}
\end{align}
with $p(t,n)$ given by Eq.~\eqref{eq:p_soln}. In particular, for the case $g<l$, by summing the series in $n$ explicitly, we get
\begin{align}
        \bar\omega_T
        = \frac{1}{l-g} + \frac{2l}{\pi T}
        \int_{0}^{\pi} \frac{\lambda(g,l,\theta)(1 - {\lambda(g,l,\theta)}^T) \sin^2\theta}
        { {(1 - \lambda(g,l,\theta))}^2 } d\theta.\label{wteqn}
\end{align}
\begin{figure}[tbh]
\centering
\includegraphics[width=8cm]{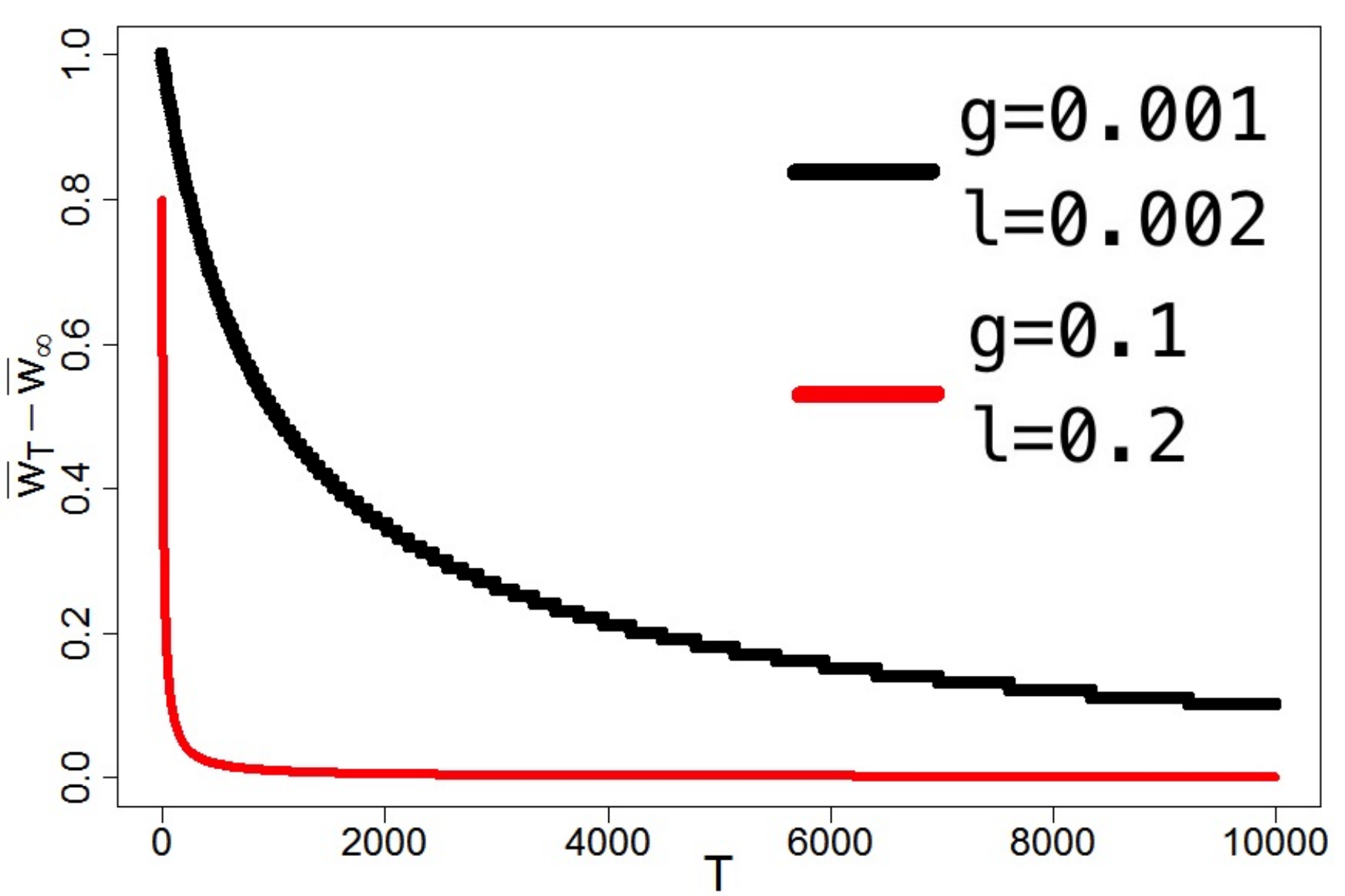}
\caption{The average waiting time of $\mathcal G$ agents as a function of discrete time in two cases; black plot: $g=0.001, l=0.002$; red plot: $g=0.1, l=0.2$.}
\label{wt}
\end{figure}

In Fig.(\ref{wt}) we plotted Eq.(\ref{wteqn}) for different values of $g$ and $l$, in the regime that $l>g$. As one can see clearly, for small $g$ and $l$, we have slow convergence of $\bar\omega_T$ to its equilibrium value. This will be reflected in the numerical simulations in Sec.~\ref{numerics}, and could be important for certain logistics systems in actual practice.

\subsection{The Non-interacting Limit}\label{limit}
The analysis above is completely general, as any system can be rescaled to satisfy $0<l+g<1$ by redefining the time step interval $\Delta t$. However, either physically or numerically, we can also encounter situations in which at smallest possible $\Delta t$ both $l$ and $g$ can be quite large (as we would see in Sec.~\ref{numerics}). In such cases, the dynamics of the system can be understood more straightforwardly in the non-interacting limit.

The analysis is non-trivial when $0<g+l<1$ because queuing of $\mathcal G$ agents will invariably occur at some, if not most, of the time steps. Heuristically speaking, queuing $\mathcal G$ agents interact with each other, as agents at the back of the queue will not see an effective pickup rate of $l$. Such interaction requires detailed analytic treatment as we presented in Sec.~\ref{snm}.

Let us now look at a single $\mathcal G$ agent generated at the node. Assuming no further $\mathcal G$ agents are generated, the probability of this agent surviving for exactly $n$ time steps is given by $p_n=\left(1-l\right)^nl$. The average waiting time of this agent is thus given by
\begin{eqnarray}\label{single}
\bar \omega=\sum_{n=0}^\infty np_n=\frac{1}{l}-1
\end{eqnarray}
Eq.(\ref{single}) is obviously only valid for $l\leq 1$, and in the case that $\mathcal G$ agents are continuously generated with probability $g$ at each time step, the average waiting time for all $\mathcal G$ agent is no longer given by Eq.(\ref{single}). This is because when there is queuing at the node, the effective probability of the generation of $\mathcal L$ agent, $l'$, for the queued agents are reduced, due to this rather non-trivial interaction between existing $\mathcal G$ agents at the node.

The non-interacting limit can be realized with $l=1$ and $g<1$. In this case, each time step there is always one $\mathcal L$ agent at the node. Given that $g<1$, $\mathcal G$ agents will not interact with each other as there will be no queuing. In principle, for any $l$ we can rescale $\Delta t$, the time interval between two consecutive time steps, to $\Delta t'=\Delta t/l$. We thus map to the case of $l=1$, Eq. (\ref{single}) can be applied for all $\mathcal G$ agent in the system, with the average waiting time as \emph{``zero"}. Note that physically this means the average waiting time is smaller than the simulation time step $\Delta t'$, which will not be resolved in numerical simulations. Thus Eq.(\ref{single}) is only useful for large rate of generation of $\mathcal L$ agents; otherwise for very small rate, $\Delta t'$ needs to be large for the non-interacting limit to apply, and the conclusions from the non-interacting limit is not very useful. For example, if $\Delta t'$ is one hour, only knowing that the average waiting time is less than one hour from the non-interacting limit cannot help us optimise it.

This non-interacting limit is thus usually the case for realistic logistics systems where $l$ is generally large even with small time resolution. The dynamics of such systems can be more straightforwardly understood in the formalism presented here. The simplification comes from the fact that the detailed dynamics of the system within the time resolution $\Delta t$ are ignored. Such dynamics are still fully captured in the more general analysis in Sec.~\ref{snm}, though they may not be physically relevant when the non-interacting limit is applicable. As we will see in Sec.~\ref{numerics}, for taxi systems with booking policies, the non-interacting limit generally applies. In contrast, with stochastic policies corresponding to road-side hailing, the more general analysis is needed to explain certain features of the taxi dynamics.

\subsection{The Network Effects}

The generalisation of the single node mode to the entire network is straightforward if we know the spatial distribution of the rate of generation of $\mathcal L $ agent, $l_i$ (where $i$ is the node index). It is, however, rather non-trivial to determine this distribution, which depends on many factors of the transport system. There are two parts contributing to $l_i$: one part comes from the intrinsic generation of new $\mathcal L$ agents, the other part comes from moving of $\mathcal L$ agents from one node to another. For the latter, it depends both on the underlying network and the policy matrix governing the movements of $\mathcal L$ agents. Microscopically, the equation governing the dynamics of $l_i$ is given as follows:
\begin{eqnarray}
l_{i,t+1}&=&\mathcal P_{ii}\left(l_{i,t}-\bar s_{i,t}\right)+\nonumber\\
&&\sum_j\mathcal P_{ji}\left(l_{j,t-\mathcal A_{ji}^{-1}}-\bar s_{j,t-\mathcal A_{ji}^{-1}}\right)+s_{i,t}\label{rwalk}
\end{eqnarray}
Using the taxi system as an example, here $l_{i,t}$ can be viewed as the number of empty taxis at the $i^{\text{th}}$ node at time $t$. The policy matrix $\mathcal P_{ij}$ and the adjacency matrix $\mathcal A_{ij}$ are both given and static; $s_{i,t}$ is the source term describing the intrinsic generation of new $\mathcal L$ agents at the beginning of the time step, while $\bar s_{i,t}$ is the sink term describing the intrinsic removal of existing $\mathcal L$ agents at the end of the time step. Both $s_{i,t}$ and $\bar s_{i,t}$ depends on the detailed dynamics of the logistics system. Thus in Eq.(\ref{rwalk}), the first line gives the number of remaining $\mathcal L$ agents at the $i^{\text{th}}$ node at the beginning of the time step (that survives the sink term), who chooses not to move to neighbouring nodes. The first term in the second lines gives the number of $\mathcal L$ agents moving from neighbouring nodes. Eq.(\ref{rwalk}) is not equivalent to diffusion or random walk on graphs\cite{graph} where the edge weights are interpreted as transition probabilities from one node to another. Here, the agents have to travel from one node to another, and the edge weights are inversely proportional to the time it takes for the $\mathcal L$ agent to move between neighbouring nodes. While on average we would expect Eq.(\ref{rwalk}) can be effectively described as some diffusion processes, exactly solving Eq.(\ref{rwalk}) is difficult. It does allow us to numerically study the time evolution of $l_i$ without doing the full scale agent based modelling, which we will explore in Sec.~\ref{numerics}.

\section{Microscopic Numerical Simulations}\label{numerics}

The treatment of the logistics system in the previous section is highly simplified, and such simplifications are necessary for obtaining analytical results applicable to very large system sizes that could be too resource intensive for numerical simulations. Our goal, however, is always trying to understand the dynamical behaviours of realistic logistics systems, with many microscopic details that are ignored in the previous section. In this section, we carry out large scale agent based numerical simulations of more realistic systems, to understand their dynamics based on the theories developed in the previous section, as well as to validate the analytical results we have obtained. The comparison of the numerical results with the results from the simplified models in Sec.~\ref{theory} shows those models indeed captures the most useful part of the dynamics of the complex realistic systems.  

Our large scale simulation platform consists of (1) an underlying network, (2) the agents of $\mathcal G$ with their origins and destinations denoted by the nodes in the network, and (3) the agents of $\mathcal L$ that move along the edges of the network from one node to another. The positions and status of all agents are updated at every time step, which we can nominally represent as one second. The agents $\mathcal G$ are generated stochastically according to $g_i$; thus at each time step, a ``dice" is thrown for each of the nodes to determine if a new agent is generated at this particular node. The generation of agents $\mathcal L$ can be either done similarly with $l_i$, or via any specific ``hidden dynamics" that we can now fully capture with microscopic simulations. For example, an agent $\mathcal L$ can roam the network according to some predetermined policy matrix $\mathcal P_{ij}$, bind with an agent $\mathcal G$ if they happen to be at the same node at a specific time step, delivering it to its destination node and become a free agent again. The action items in a single time step is illustrated in Fig.(\ref{schematics}), which is looped over repeatedly for the entire simulation.
\begin{figure}[tbh]
\centering
\includegraphics[width=18cm]{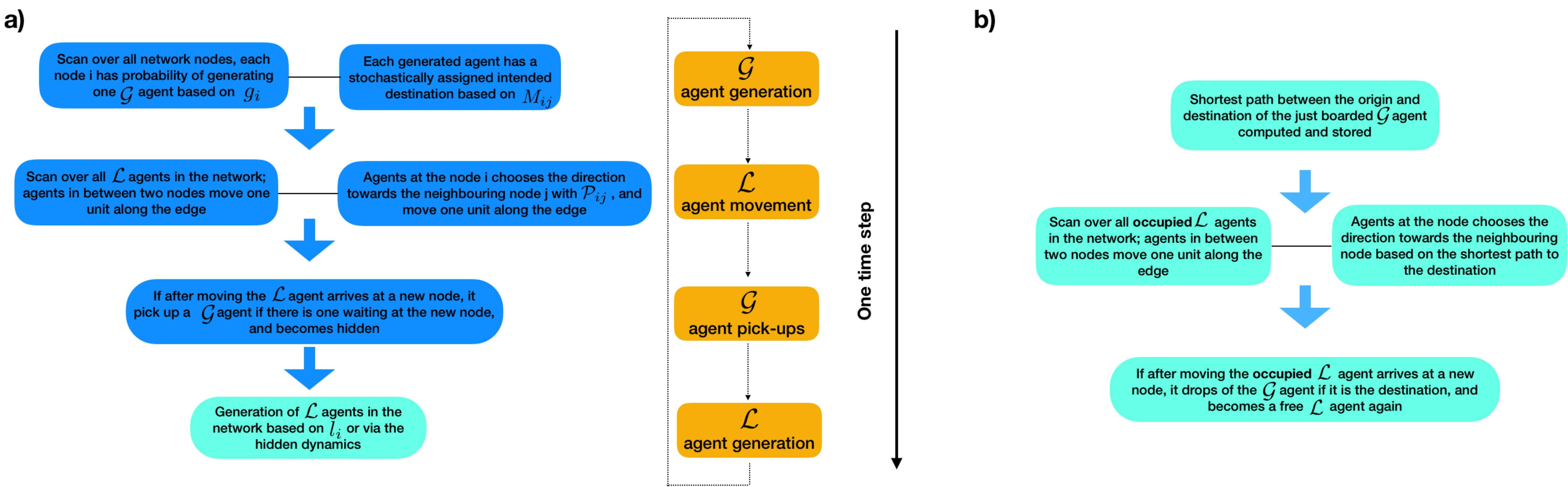}
\caption{Action items in a single time step both for $\mathcal G$ agents and $\mathcal L$ agents. Panel a) illustrates the dynamics captured in the theoretical analysis in Sec.~\ref{theory}. Panel b). illustrates the ``hidden dynamics" not captured in Sec.~\ref{theory}, but captured in agent-based simulations.}
\label{schematics}
\end{figure}

As a concrete example, we carry out detailed microscopic simulation of the taxi systems as described in Sec. ~\ref{ps}, and compare the simulation results with the analysis in Sec.~\ref{theory}. We focus on the cases where the road network $\mathcal A_{ij}$ and the commuter generation probability distribution $g_i$ are both time-independent. The empty taxi generation probability $l_i$ is time dependent and governed by the dynamics of taxis picking up commuters and dropping them off at the destinations, in addition to empty taxis roaming about the road network. We thus require the input of an origin/destination matrix $M_{ij}$ and the policy matrix of the empty taxis $\mathcal P_{ij}$ as introduced in Sec.~\ref{ps}, which we also take to be time independent. For each simulation, the total number of taxis (the sum of empty taxis and occupied taxis) is conserved. We would like to emphasize that for occupied taxis (which is not treated as agents in our theoretical formulation), they move from the origin of the boarded commuter to the respective destination via the shortest path computed from the road network using the Dijkstra's algorithm.

\subsection{Stochastic Policies and Booking Policies}

Given a specific road network and the commuter demand patterns (as encoded in $g_i$), we would expect the dynamics of the taxi system to be strongly dependent on the movement of the empty taxis governed by $\mathcal P_{ij}$. Two types of the policies are considered in this work. The first type is the completely decentralised stochastic policies, where $\mathcal P_{ij}$ is the same for every empty taxi looking for commuters. This type corresponds to the traditional road-side hailing of taxis, in which empty taxis do not know where exactly commuters are waiting. At each node, $\mathcal P_{ij}$ gives the probability distribution of which neighbouring node the empty taxi will move to. The distribution itself can be random or highly non-trivial. Instead of being completely random, experienced taxi drivers know where to look for commuters with a more effective $\mathcal P_{ij}$. While we do not empirical data on real driver behaviours, we expect the empirical policy could be more efficient, leading to reduction of number of taxis needed and reduced waiting time. The qualitative dynamical properties revealed in this work, however, should remain the same.

The second type is based on real-time booking of taxis, nowadays quite common via smartphone Apps or telephone calls. In this case, when a commuter is generated at any node in the network, a certain algorithm is implemented to assign a nearby empty taxi to come pick the commuter up. The assigned empty taxi becomes booked, which will go from its current position to the location of the commuter via the shortest path. In this way, the policy matrix is different for different booked taxis (which are still counted as agents of type $\mathcal L$), which we formally denote as $\mathcal P^m_{ij}$ (the superscript is the taxi index). In this work, we use a simple assignment algorithm, in which the nearest empty taxi within a specific range $R$ from the commuter will be assigned. For empty taxis that are not booked, they still move with a decentralized policy $\mathcal P_{ij}$, and in our simulations it is a completely random policy. Thus empty taxis roam the road network with a random walk, before being booked and move directly towards the origin of the booking commuter.

\subsection{Spatial Generation of Empty Taxis}

The spatiotemporal generation of empty taxis, denoted as $l_i\left(t\right)$, is highly important from resource allocation point of view. Given that the total number of taxis is fixed, $l_i\left(t\right)$ depends strongly on $g_i, M_{ij}$, the road network structure $\mathcal A_{ij}$, and the policy matrix $\mathcal P_{ij}$. An accurate prediction of the empty taxi generation pattern is difficult to solve analytically from Eq.(\ref{rwalk}); we thus employ a practical approach for most of the cases, and study the pattern of $l_i$ numerically. Decentralized movement (i.e. road-side hailing) and taxi booking will be considered separately; and since we only look at cases where $g_i, M_{ij}, \mathcal A_{ij}, \mathcal P_{ij}$ are static, we will only focus on the spatial distribution of $l_i$ after the system reaches equilibrium. In Fig.(\ref{schematics2}) we show a schematic drawing of the artificial road network, how taxis disappears from the network from picking up commuters, as well as reappear after dropping off commuters at their destinations.
\begin{figure}[H]
\centering
\includegraphics[width=12cm]{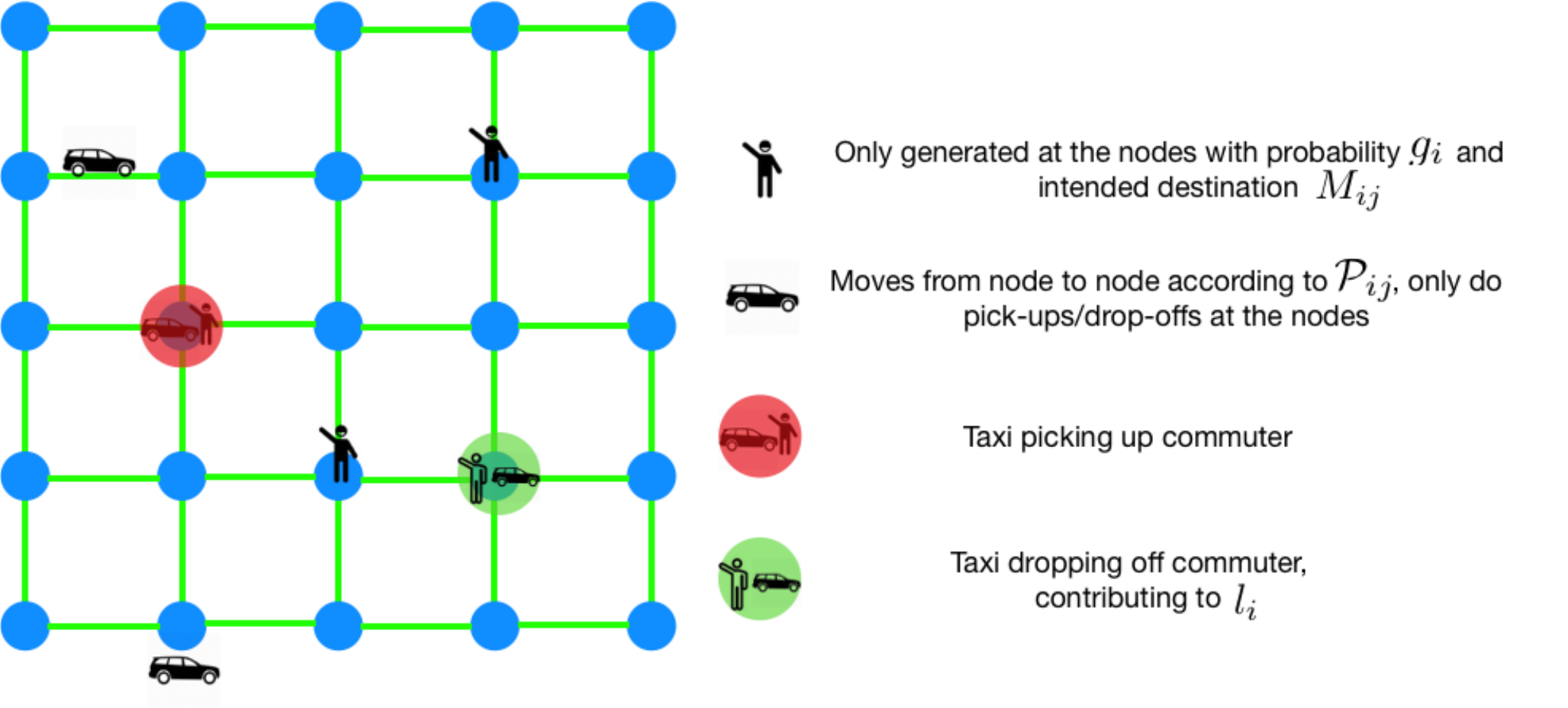}
\caption{Schematic representations of the road network and the two types of agents in the network.}
\label{schematics2}
\end{figure}

For the case of road-side hailing, $l_i$ has two contributions at the $i^{\text{th}}$ node: $d_i$, the empty taxi generated when an occupied taxi delivered a commuter at its destination at node $i$; $m_i$, coming from an empty taxi moving to node $i$ from a neighbouring node. In particular, $d_i$ is equivalent to the source term in Eq.(\ref{rwalk}). In the special case where the destination is uniformly distributed, so that $M_{ij}=1/N_{\text{node}}$, the inverse of the total number of nodes $N_{\text{node}}$, $d_i$ is uniform and independent of $i$. When the number of taxis is small so that in the long time limit all nodes have queuing commuters, $m_i$ is negligible, so that $l_i=d_i$ is also uniformly distributed. This corresponds to the case in Eq.(\ref{rwalk}) where $l_{i,t}=\bar s_{i,t}$ at any time step. When the number of taxis is large and $m_i$ is sufficiently large, both $m_i$ and $l_i=d_i+m_i$ are only uniform if there is full translational symmetry of the road network and the policy matrix: $g_i=g$ is constant and independent of $i$, $\mathcal A_{ij}$ gives a regular lattice with all edges of the same weight, and $\mathcal P_{ij}=1/n_i$, where $n_i$ is the number of neighbours of the $i^{\text{th}}$ node. In particular, if $\mathcal A_{ij}$ is an arbitrary network, $m_i$ does not generally equilibrate as one can see in Fig. (\ref{inflow}).

\begin{figure}[tbh]
\centering
\includegraphics[width=16cm]{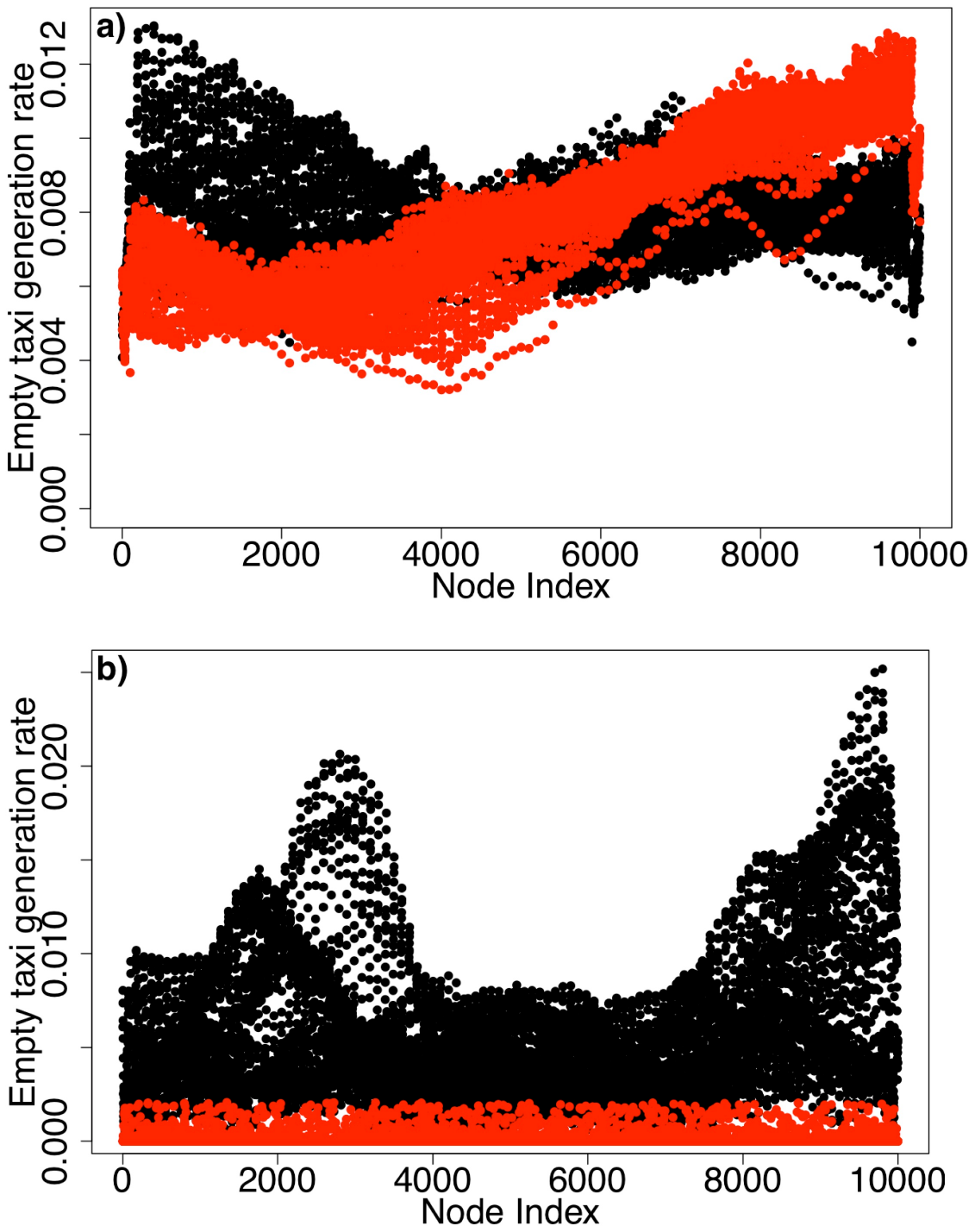}
\caption{a). Numerical simulation of $l_i$ at each node with uniform distribution given by $g_i=g$ and $M_{ij}=1/n_i$. The black and red plots are two different simulations with randomly distributed empty taxis as an initial condition, with a square lattice network with non-uniform edge weights of a total 10000 nodes. The simulations are done with 300 taxis, and different simulations give different spatial distribution of $l_i$; b). The same square lattice network with the same number of taxis, but with non-trivial $g_i$ and $M_{ij}$. The spatial distribution of $l_i$ (black plot) is the same for repeated simulations. The blue plot gives the spatial distribution of $d_i$. }
\label{inflow}
\end{figure}

In the case that $M_{ij}$ is non-uniform, on the other hand, $l_i$ is dominated by the spatial distribution of $M_{ij}$, and $\mathcal A_{ij}$ only has a sub-leading effect. It is useful to look at the dependence of $l_i$ on $g_i$, which we plotted in Fig.(\ref{inflow2}). While $d_i$ depends on $M_{ij}$ only, $m_i$ also strongly depends on $\mathcal P_{ij}$, in addition to $\mathcal A_{ij}$. In particular, from a heuristic point of view a good $\mathcal P_{ij}$ will lead to higher $l_i$ if $g_i$ is larger: if the node has a higher probability of generating commuters per time step, the effective arrival rate of empty taxis at this node should also be higher. In Fig.(\ref{inflow2}a) the two plots are from a random walk implemented with $\mathcal P_{ij}=1/n_i$, and a more intelligent $\mathcal P_{ij}$ from the recursive value model enhanced with reinforcement learning\cite{yangli}. One can clearly see for the latter, $l_i$ increases appreciably with $g_i$, indicating more efficient policy matrix for commuter seeking.

With taxi-booking, we look at the case where the commitment of booking is binding: an empty taxi that is booked can only pick up the respective commuter, and a commuter who books a taxi will only board the assigned taxi. Note also for empty taxis that are not booked yet, they just roam randomly in the road network. In this way, $l_i$ only has contribution from the booked empty taxis arriving at the locations of their respective commuters. The approximately linear relationship between $l_i$ and $g_i$ is evident from the numerical simulations, as one can see from Fig.(\ref{inflow2}b). When the taxi number is small, almost all taxis are booked immediately after they become empty (finish delivering the passengers). This is way there is no random movement at all. The inflow of booked taxis is proportional to the rate of the generation of passengers at each node, resulting in an almost straight line (see the black plot in Fig.(\ref{inflow2}b)). One should also note that in contrast to the stochastic policies where the interaction of the commuters and empty taxis are strictly local (confining to the same node), with taxi-booking such interaction is intrinsically non-local, depending on the range of booking $R$. The range of booking is defined as follows: only empty taxis within a radius of $R$ from the commuter has a chance to be booked. We will discuss about such distinction about the locality of interaction in Sec.~\ref{dphase}.
\begin{figure}[tbh]
\centering
\includegraphics[width=16cm]{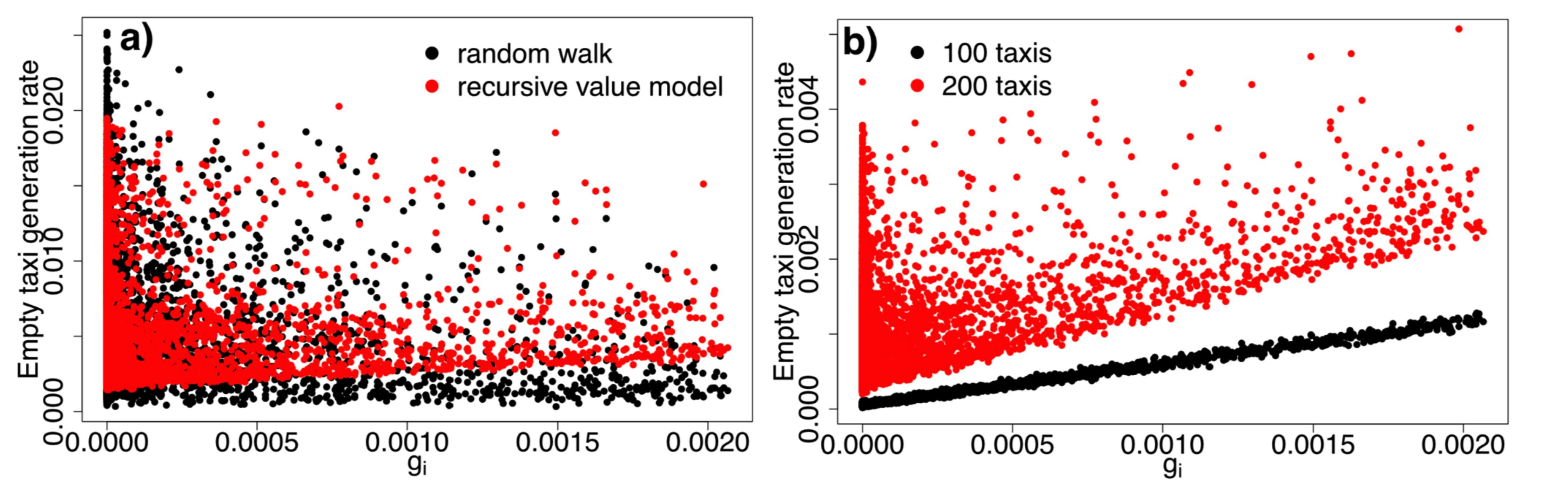}
\caption{a) The numerical simulation of $l_i$ with taxi-booking for 100 taxis, where $l_i$ is given by rate of booked taxis arriving at corresponding commuters' location at $i^{\text{th}}$ node; b). The numerical simulation of $l_i$ with taxi-booking, where $l_i$ is not only given by rate of booked taxis arriving at corresponding commuters' location at $i^{\text{th}}$ node, but also by empty taxis passing by nodes or occupied taxis delivering passengers at the node. Note for the latter, these empty taxis are not capable of picking up commuters if they are not booked.}
\label{inflow2}
\end{figure}

\subsection{Critical Taxi Number $N^*$}

Since each node in the network has a well-defined rate of generation of $g$ (for commuters) and $l$ (for empty taxis), we can use the results from the models in Sec.~\ref{theory} to analyze if there will be queuing for each node, and the overall average waiting time of the commuters of the entire system. In particular, from Sec.~\ref{theory} we know that for each node there is a second-order phase transition at $g=l$. Heuristically speaking, in the long run there will be accumulation of commuters at the node if $g>l$, and no accumulation if $g\leq l$. For the entire road network, there will be no commuter accumulation if $g_i<l_i$ for each node; otherwise if we assume commuters will wait for taxis no matter how long it takes, the average waiting time of the commuters will diverge. This generic argument should apply to a wide variety of taxi systems, including various forms of road-side hailing and taxi-booking. One should note that in this case, average waiting time of the commuters is equivalent to the average number of commuters in the road network, over all the time steps. Formally, the average waiting time is given by
\begin{eqnarray}
\bar w_T=\frac{1}{Tg}\sum_{t=1}^{T}N_{\mathcal G}\left(t\right)
\end{eqnarray}
where $g=\sum_i g_i$, $T$ is the total number of time steps for simulation, and $N_{\mathcal G}\left(t\right)$ is the number of commuters (the $\mathcal G$ agents) in the entire road network at time step $t$. For a single node it reduces to Eq.(\ref{singlenode}).

Given the complicated dependence of $l_i$ on various aspects of the taxi system (especially the details of the road network and the policy matrix), we explore numerically how the average waiting time of the taxi system depends on the number of taxis $N_{\text{taxi}}$ (sum of empty and occupied taxis) in the simulation. We expect $l_i$ should increase with $N_{\text{taxi}}$ if all other aspects are fixed. Thus for small $N_{\text{taxi}}$, we expect $g_i>l_i$ for majority of nodes, while for large $N_{\text{taxi}}$ we have $g_i<l_i$ for majority of the nodes.

\begin{figure}[tbh]
\centering
\includegraphics[width=9cm]{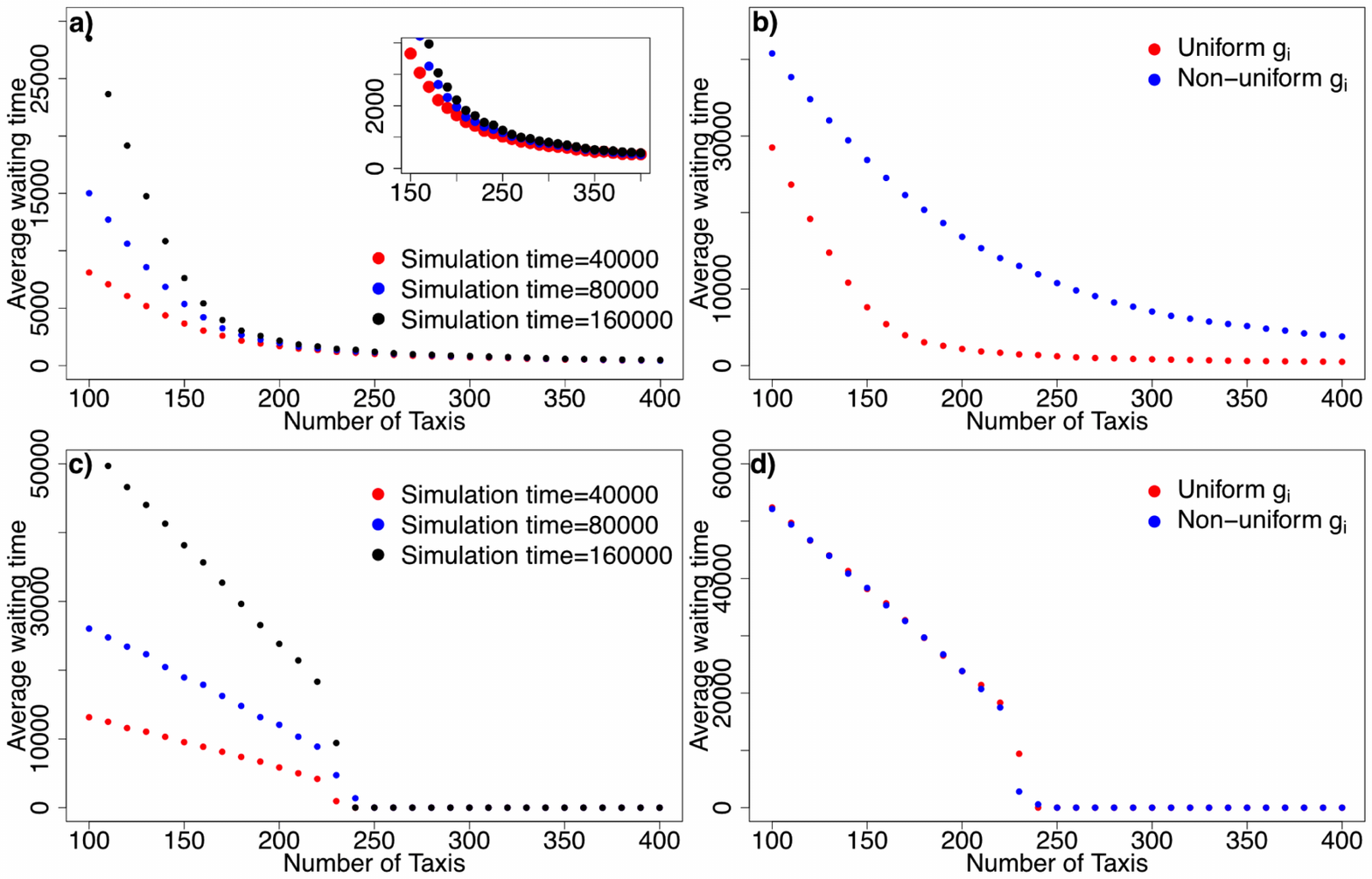}
\caption{The average waiting time of commuters generated in the simulation, as a function of the total number of taxis in the system. a). Simulation results from a square matrix of non-uniform edges and a total of $10000$ nodes, and with uniform $g_i$, with plots of different total simulation time and a random walk policy $\mathcal P_{ij}$; b). Comparison of the same road network but with different spatial distributions of commuters, with a random walk policy $\mathcal P_{ij}$; c) Same as a) but with a booking policy with infinite range; d). Same with b) but with a booking policy with infinite range. }
\label{phase}
\end{figure}

For stochastic policies, we let empty taxis roam randomly in the road network. Different stochastic policies lead to qualitatively similar results, which we will discuss in details elsewhere\cite{yangunpublished}. It is clear from the numerical simulation that when the number of taxis $N_{\text{taxi}}$ is large enough, given a spatially uniform rate of generation (here we look at the simple case of $g_i=1/N_{\text{node}}$, where $N_{\text{node}}$ is the total number of nodes of the network, and one commuter is generated in the entire road network per time step), eventually $l_i>g_i$ for every node $i$. In this regime, the average waiting time no longer grows with the simulation time, as there is no queuing of commuters at any node. Conversely, as $l_i$ decreases with decreasing $N_{\text{taxi}}$, queuing will eventually occur at nodes with $g_i>l_i$. In this case, the average waiting time will increase with $T$, as one can see in Fig.(\ref{phase}). If both $l_i$ and $g_i$ are uniform, a sharp transition can be observed in the limit of large $T$, as one can see the trend in Fig.(\ref{phase} a). For non-uniform $l_i$ and $g_i$ and finite $T$, there will be a much smoother transition area in which queuing only occurs at some nodes. Thus for actual finite $T$ simulations, the case with non-uniform $g_i$ will lead to a much smoother dependence of $\bar w_T$ on $N_{\text{taxi}}$, as one can see in Fig.(\ref{phase} b).

While the dynamics of the stochastic policies can be generally understood via the analysis in Sec.~\ref{snm}, the booking policies in most cases correspond to the non-interacting limit as we discussed in Sec.~\ref{limit}. The general idea of the booking policy is that when a commuter is generated, an optimal taxi is chosen and given the information about the location of this commuters. This taxi then becomes booked and heads directly towards the respective commuter. In contrast to the stochastic policies where the interaction between taxis and commuters is localized (the empty taxi has no information about the location of the waiting commuters until it is at the same node of the waiting commuters), such interaction is global with the booking policies, when the information of the location of the waiting commuter is made available to the booked empty taxi that potentially can be quite far away, depending on the range of the booking policy, $R$.

To understand that, let us look at the simple example, where the nearest empty taxi (if available) at a particular time step is booked for each of all commuters generated at that time step, no matter how far away this empty taxi is (i.e. $R\rightarrow\infty$). The trip time for the booked taxi to arrive at the waiting commuter is bounded given the finite size of the network; its statistical average is well-defined and can be computed from the road network and the density of the empty taxis. Apart from this contribution to the average waiting time, we can thus treat the entire network as a single ``node", with the rate of generation of commuters and empty taxis given by $g=\sum_i g_i$ and $l=\sum_il_i$ respectively. The non-interacting limit thus applies if $g\ge 1$ (since in all simulations we take $\Delta t$ as one second between each time step), which is generally the case for any realistically large road network. In Fig.(\ref{phase} c,d), we take $g=1$. We thus observe a much sharper transition even for small $T$ in Fig.(\ref{phase} c), and the dynamics is not affected by the distribution of $g_i$ because of the global nature of the booking policies. The transition occurs at $l=g$, and when $l>g$ there is no queuing at any nodes (every commuter generated can book an empty taxi almost right away). Such qualitative behaviours should apply to other booking policies, i.e. the more realistic cases with finite $R$.

There is thus a well-defined critical number of taxis $N^*_{\text{taxi}}$ in the limit of $T\rightarrow\infty$, such that when $N_{\text{taxi}}>N^*_{\text{taxi}}$ there are no queuing at any node, and when $N_{\text{taxi}}<N^*_{\text{taxi}}$ there are queuing of commuters at some or all of the nodes in the road network. For the booking policies with $R\rightarrow\infty$, $N^*_{\text{taxi}}$ occurs when $g=l$, and a rather sharp transition occurs at $N^*_{\text{taxi}}$ as one can see from Fig.(\ref{phase}c,d), where $N^*_{\text{taxi}}\simeq 245$. For booking policies with finite range $R$, or with stochastic policies, $N^*_{\text{taxi}}$ needs to be large enough so that for $N_{\text{taxi}}>N^*_{\text{taxi}}$, $l_i>g_i$ for all $i$. For finite $T$, the transition at $N^*_{\text{taxi}}$ may not be sharp because of the slow convergence of the taxi dynamics as one can see from Eq.(\ref{wteqn}) and Fig.(\ref{wt}). When $g_i$ or $l_i$ are non-uniform, we also expect a certain range of $N_{\text{taxi}}$ in which only a portion of nodes have queuing commuters. Nevertheless, the transition will be sharp in the limit $T\rightarrow\infty$, and $N^*_{\text{taxi}}$ is still well-defined. In particular, for stochastic policies with random walk, $N^*_{\text{taxi}}\simeq 195$ as one can see from Fig.(\ref{phase}a), with the same road network and the same total rate of generation $\sum g_i=1$.

Interestingly, $N^*_{\text{taxi}}$ is smaller in the case of the stochastic policies, as compared to the booking policy with no range constraints. In general, $N^*_{\text{taxi}}$ quantitatively depends on all details of the taxi system, and it is an important quantity characterising the efficiency of the taxi system. We also expect this critical number of delivering agent to be a universal concept the dynamics of the bi-agent logistics system in an arbitrary network. It serves as the boundary of two distinctive dynamical phases of such logistics system, as we will illustrate below in Sec.~\ref{dphase}.

\section{Dynamical Phases of the Taxi System}\label{dphase}

The qualitatively different dependence of the average waiting time on the simulation time for $N_{\text{taxi}}<N^*_{\text{taxi}}$ and $N_{\text{taxi}}>N^*_{\text{taxi}}$ allows us to define two distinct dynamical phases of the taxi system, with $N^*_{\text{taxi}}$ serving as the phase boundary. We denote the phase with $N_{\text{taxi}}<N^*_{\text{taxi}}$ the \emph{oversaturated phase}, where the demand for taxis exceeds the supply, leading to queuing of commuters at some or all of the nodes in the road network. On the other hand, the phase with $N_{\text{taxi}}>N^*_{\text{taxi}}$ is denoted as the \emph{undersaturated phase}, where the supply of taxis exceeds the demand. In this case, no nodes have queuing commuters in the long run, and there is an excess of empty taxis running around looking for commuters. From the logistics point of view, given a specific demand and a specific road network, a supply of $N^*_{\text{taxi}}$ is optimal, in the sense that all taxis are efficiently utilised, and all commuters can be picked up within reasonable amount of waiting time. When the supply of taxis decreases from $N^*_{\text{taxi}}$, the average waiting time of the commuters increases rapidly, suggesting a severe degrading of the service quality of the taxi system. In contrast, when the supply of taxis increases from $N^*_{\text{taxi}}$, the average waiting time of the commuters only decreases marginally. On the other hand, the cost of supplying taxis increases linearly (in the form of fuel cost and/or compensation to drivers). It is thus undesirable to deviate too much from $N^*_{\text{taxi}}$ on both sides, and it is important to know that given a particular taxi system, what is the optimal number of taxis $N^*_{\text{taxi}}$ of this system.

{\it Implications in optimisation--} Once we have as many details as possible (e.g. traffic conditions, commuter loading and unloading time, etc.), comprehensive numerical simulation can lead to accurate prediction of $N_{\text{taxi}}^*$ given a specific demand, as well as the corresponding average waiting time $\bar w^*$ at $N^*_{\text{taxi}}$. This has important implications on how we should optimise the average waiting time. While conventionally, the optimal number of taxis in a city is determined by making sure that the average waiting time is below a certain chosen benchmark $\bar w_{\text{bm}}$ (e.g. $\bar w_{\text{bm}}=5\text{min}$). We now know that if $\bar w_{\text{bm}}>\bar w^*$, we can easily increase the number of taxis in the oversaturated phase to reduce the average waiting time to $\bar w_{\text{bm}}$. On the other hand, if $\bar w_{\text{bm}}<\bar w^*$, we need to increase the number of taxis dramatically in the undersaturated phase to achieve $\bar w_{\text{bm}}$, which may not be the best option. In the latter case, we should focus on improving the empty taxi routing policies or booking algorithms, or other approaches to shift $N^*_{\text{taxi}}$, which could be a more economical way in reducing the average waiting time, in contrast to the increase in the total number of taxis in a city.

\subsection{Tuning of $N^*_{\text{taxi}}$}\label{tuningn}

The critical taxi number $N^*_{\text{taxi}}$ most directly depends on the rate of generation of commuters $g_i$. When the total demand increases, more taxis are needed if all other factors are kept constant. With the same total demand given by $g=\sum_ig_i$, different spatial distribution (given by the dependence of $g_i$ on $i$, the node index), the destination distribution (given by the OD matrix $M_{ij}$), the underlying road network (given by the adjacency matrix $A_{ij}$), and the manoeuvring strategies of empty taxis (given by the policy matrix $P_{ij}$)  can also lead to variation of $N^*_{\text{taxi}}$. In most cases, the quantitative analysis can only be done numerically, as the effective $l_i$ strongly depends on those factors in highly non-trivial ways.

In this part we analyse the dependence of the dimensionless regularized waiting time $\omega_0=\lim_{N_T\rightarrow\infty}\bar\omega/T$ on the number of taxis in the system. In the undersaturated phase $N_{\text{taxi}}>N^*_{\text{taxi}}$, we have $\omega_0=0$. A second order phase transition occurs at $N^*_{\text{taxi}}$, and in the oversaturated phase with $N^*_{\text{taxi}}>N_{\text{taxi}}$, $\omega_0>0$. Such phase transition is most easily detected numerically with the booking policy. This is because as we have shown in Sec. ~\ref{limit}, equilibrium can be quickly reached for $l>1$ and $l>g$ in the non-interacting limit. Different booking policies can affect $N^*_{\text{taxi}}$, as we show here in Fig.(\ref{banalysis}).
\begin{figure}[tbh]
\centering
\includegraphics[width=7cm]{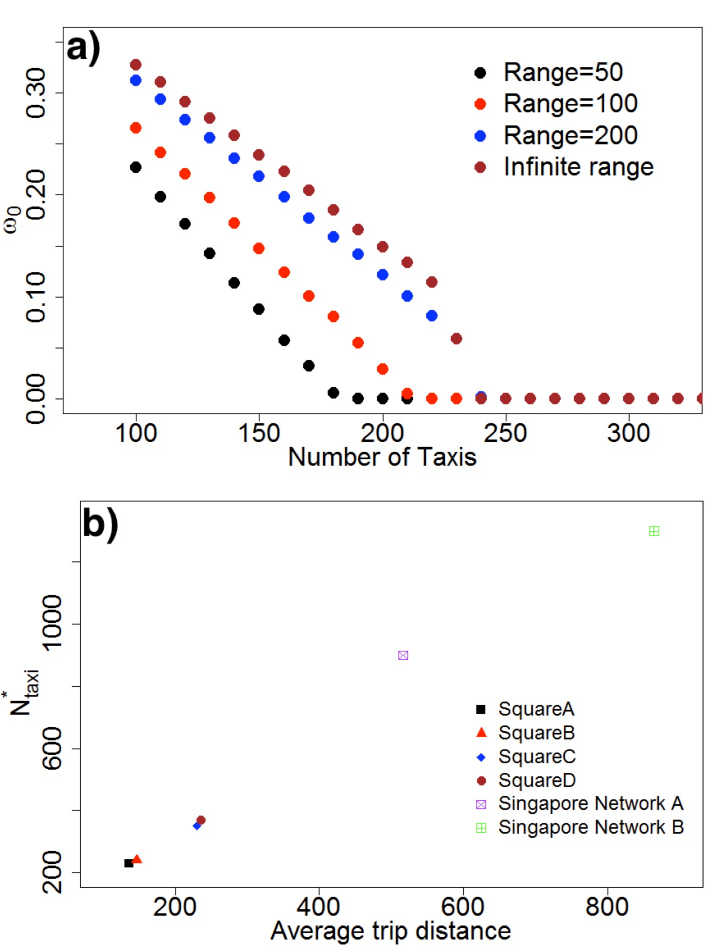}
\caption{a). Shifting of $N^*_{\text{taxi}}$ for booking algorithms with different booking range. As the booking range decreases, the critical number of taxis needed for the same demand $g_i$ and the same road network also decreases. b). The relationship between the average distances (in arbitrary unit) between origin-destination pairs generated in simulations, and the critical number of taxis $N^*_{\text{taxi}}$, for various different road networks and demand patterns (as specified by $g_i$ and $M_{ij}$). ``SquareA" to ``SquareD" are four 100 nodes by 100 nodes lattice road networks with randomly generated $\mathcal A_{ij}, g_i$ and $M_{ij}$; we also show here the empirical Singapore road network with a total of 27185 nodes. For ``Singapore Road Network A", both $g_i$ and $M_{ij}$ are synthetic and randomly generated; for ``Singapore Road Network B", $g_i$ and $M_{ij}$ are from the empirical data obtained from the actual trips taken in Singapore over one particular day.}
\label{banalysis}
\end{figure}

While all factors mentioned above can influence $N^*_{\text{taxi}}$ with the booking policies, they do not do so independently. We identify two factors that fundamentally affects the critical number of taxis, when the total average demand given by $g=\sum_ig_i$ is kept constant. The first factor is the booking policy $\mathcal P^n_{ij}$. We show in Fig.(\ref{banalysis}a) that reducing the range of booking $R$ (so that only empties within a distance of $R$ from a commuter can be booked, and the nearest empty taxi, if available, will be booked) generally reduces $N^*_{\text{taxi}}$, making the booking process more efficient in terms of the number of taxis needed for the entire taxi system. Intuitively, this is because if an empty taxi far away from the commuter is booked, it is committed and unable to take other commuters along its way to pick up the designated commuter, leading to suboptimal matching of available taxis and waiting commuters. The trade-off is that in the undersaturated phase, the average waiting time increases with decreasing $R$ (hard to distinguish from Fig.(\ref{banalysis}a) due to the scale but the change is unambiguous). Thus when there is an oversupply of empty taxis roaming on the streets, we can reduce average waiting time by increasing the range of booking, provided such alteration does not push the taxi system from the undersaturated phase to the oversaturated phase.

The other fundamental factor is the average distances between origin-destination pairs generated in the system. This factor depends collectively on $\mathcal A_{ij}, g_i$ and $M_{ij}$. The greater the average distances, the longer it takes on average for occupied taxis to send commuters from their origins to destinations, which is the hidden part of the taxi dynamics. If the total number of taxis is fixed, heuristically this will lead to lower effective $l_i$ on average, resulting in larger $N^*_{\text{taxi}}$. This is indeed the case as one can see from Fig.(\ref{banalysis}b), there is an almost linear relationship between $N^*_{\text{taxi}}$ and the average trip distance for various different cases. Both artificial lattice road network and empirical island-wide road network of the city state of Singapore are included in the plots. For the Singapore road network, we also compare the synthetic and empirical commuter demand patterns (encoded in $g_i$ and $M_{ij}$). We thus have strong numerical evidence that the linear relationship is quite general for different types of road networks and taxi demand patterns.

With booking policies, the movement of booked and occupied taxis are determined by the origins or destinations of respective commuters via the shortest path algorithm. For empty taxis that are not yet booked, their movement is completely decentralised, and in Fig.(\ref{banalysis}) we choose the simple random walk policy. More intelligent policies for empty, unbooked taxis can also be implemented; they do not alter the qualitative features illustrated in this work, and we will discuss in more details elsewhere. On the other hand, if we take the limit of the range of booking $R\rightarrow 0$, the booking policy is reduced to the stochastic policy, where all empty taxis will only pick up commuters if they are at the same node of the waiting commuters. We thus expect many of the features on the dependence of $N^*_{\text{taxi}}$ on various factors of the taxi system to be similar with the stochastic policies.

In general for realistic large road network, if physically each time step in the simulation represents one second, $g_i$ is very small for almost all nodes. Empirically for the city state of Singapore, on a typical day there are around six commuters generated for every second for the entire road network of $27185$ nodes\cite{yang}, so that $g_i\sim 0.0002$. Mathematically, $N^*_{\text{taxi}}$ is only well-defined in the limit of the simulation time $T\rightarrow\infty$, when the taxi system reaches equilibrium for $l_i$ and $g_i$ at each node. As we have shown in Sec.~\ref{theory}, convergence to asymptotic values of average waiting time can be slow even for a single node, when both $l$ and $g$ are small. Thus, theoretical investigation of the dependence of $N^*_{\text{taxi}}$ requires much longer simulation time, and extrapolation to infinite simulation time needs to be performed, for stochastic policies. Indeed, for a wide number of taxi systems, the behaviours of $N^*_{\text{taxi}}$ agrees with that with the booking policies in the limit of the booking range $R\rightarrow 0$.

\subsection{Phase characteristics in the context of ride-sharing}
\begin{figure}[tbh]
\centering
\includegraphics[width=14cm]{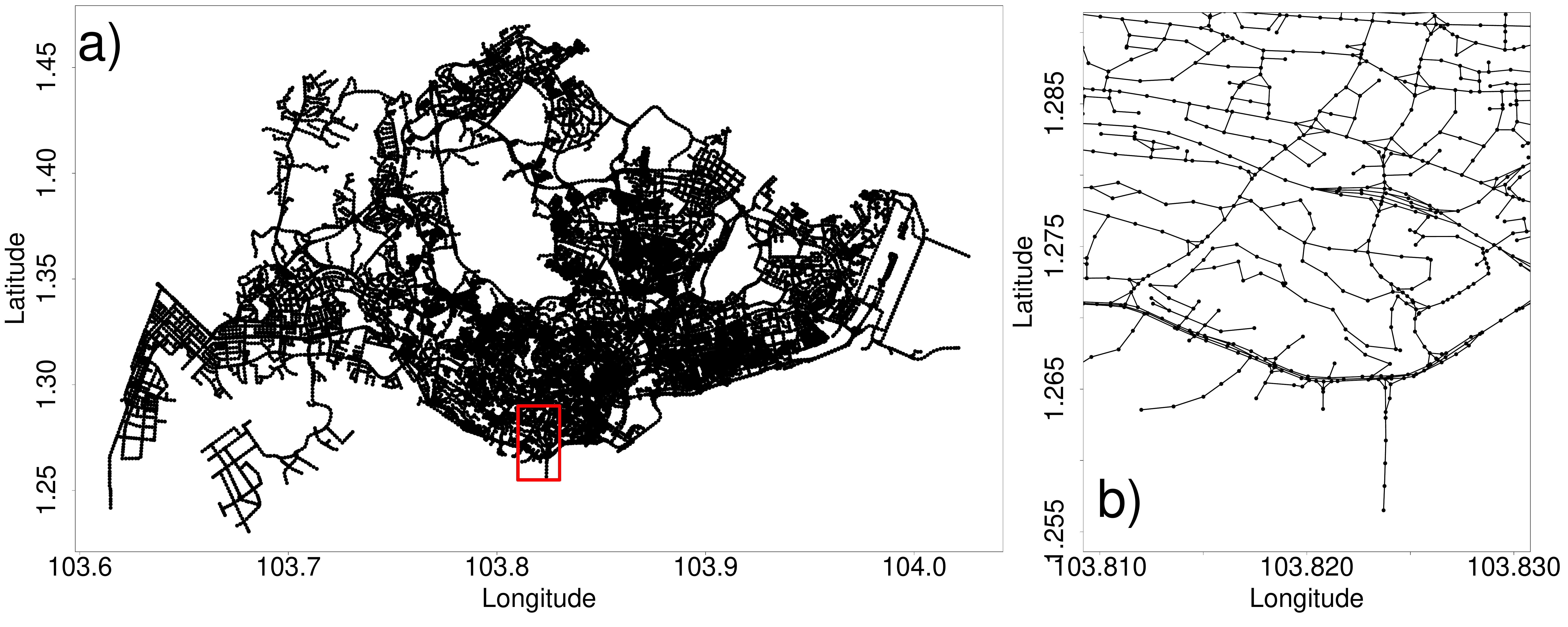}
\caption{Realistic road network of the city state of Singapore used in agent-based simulations. Panel a). is the entire road network of Singapore; Panel b) shows a zoom-in of the CBD area, within the red box in Panel a).}
\label{singaporemap}
\end{figure}
The phase transition in this bi-agent logistics system has wider applications in more complicated and realistic systems. We illustrate this with the example of the taxi systems in which real-time, dynamic ride-sharing between commuters can be implemented. In this system, if the occupant of the taxi and the waiting commuter are both willing to share their rides, and if sharing of their trips will result in detours within the acceptable limits, ride-sharing will be implemented. Strictly speaking, there are four types of active agents in this system: empty taxis, occupied taxis with with room to share and the passenger open to ride-sharing, commuters open to ride-sharing, and commuters not open to ride-sharing. For occupied taxis with no room to share or with passengers not open to ride-sharing, they are still part of the ``hidden dynamics" that is only responsible for the effective generation of other active agents. When no commuters are willing to share, the system reduces to the original bi-agent system we studied in details in this paper.
\begin{figure}[H]
\centering
\includegraphics[width=16cm]{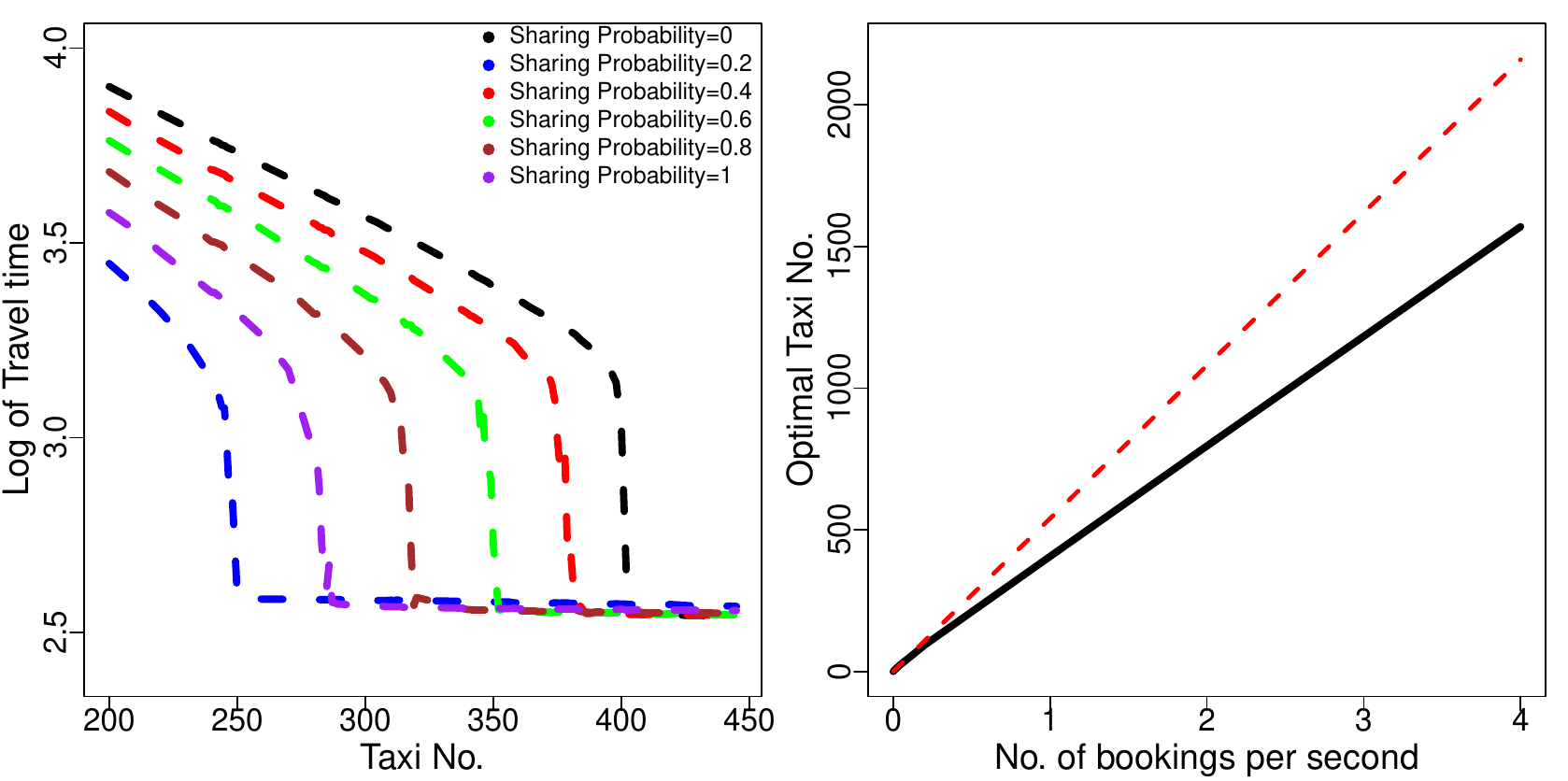}
\caption{a). The shift of the optimal taxi no. $N^*$ with different values of $p_s$. $N^*_{\text{taxi}}$ decreases monotonically but not in a smooth manner; b). The black curve is the dependence of $N^*_{\text{taxi}}$ on the taxi demand, and the red dotted curve is the straight line tangent to the black curve near the origin, showing that the dependence of $N^*_{\text{taxi}}$ on the taxi demand is sub-linear. }
\label{sharephase}
\end{figure}

Given the nature of real-time, dynamic ride-sharing, only booking policies are natural for the available taxis to pick up commuters. An important parameter to tune is $p_s$, the percentage of commuters generated in the system who are open to ride-sharing. This is a phenomenological parameter describing the degree of acceptance of ride-sharing for a particular society, and may have important cultural, social and economical implications. When $p_s>0$, the theoretical analysis in Sec.~\ref{theory} does not strictly apply. Heuristically, however, for each node there is still an effective $l$ from both empty taxis and taxis open to ride-sharing, and an effective $g$ from both types of commuters. This is confirmed numerically, and both effective $l$ and $g$ now also depend on $p_s$, the tuning of which can lead to interesting phase transitions with significant impact on the dynamics of the taxi system.

\begin{figure}[H]
\centering
\includegraphics[width=16cm]{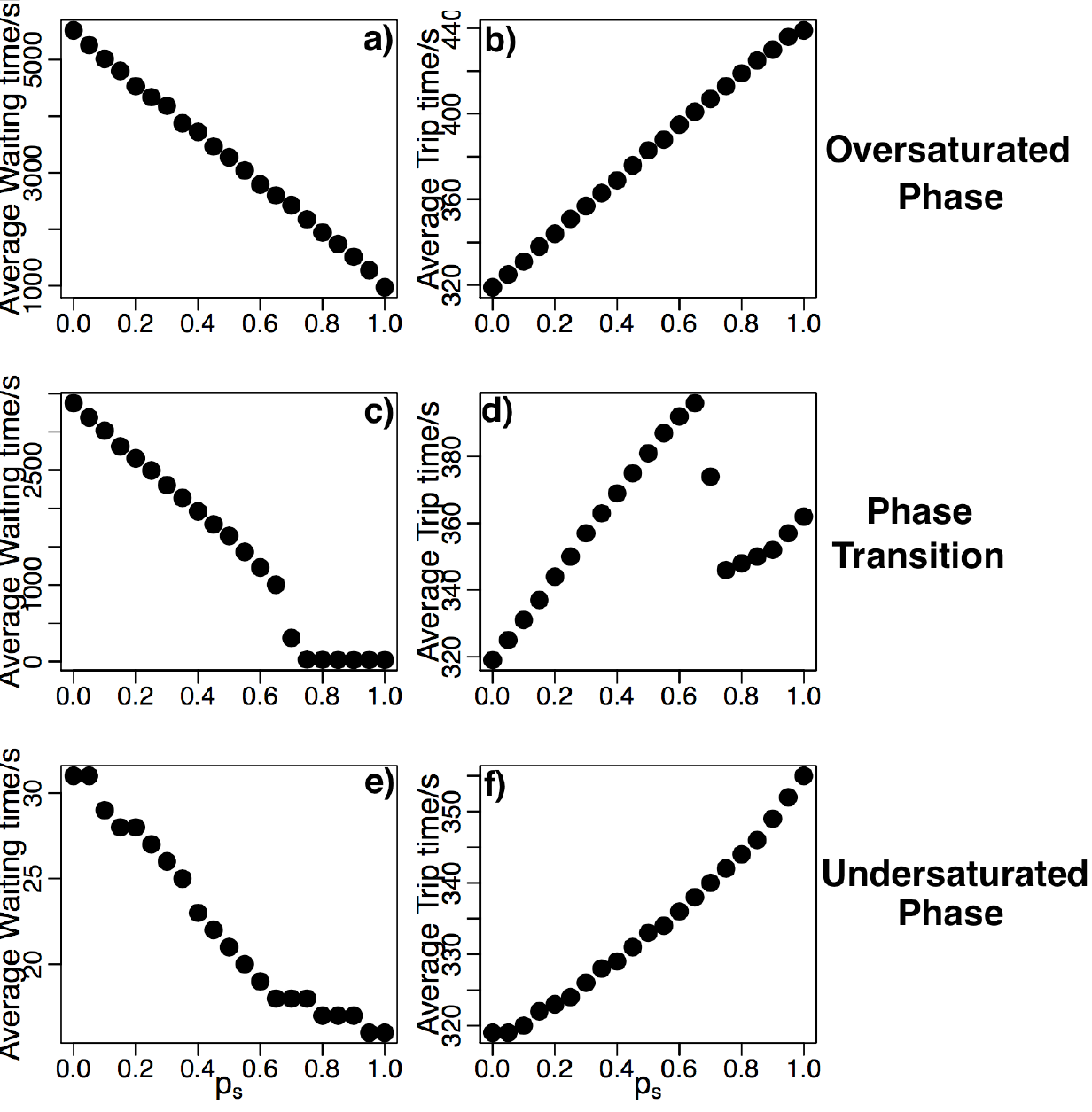}
\caption{a). Dependence of the average waiting time (black axis) and average trip time (red axis) on commuter's willingness to taxi-sharing, in the over-saturated region; b). Dependence of the average waiting time (black axis) and average trip time (red axis) on commuter's willingness to taxi-sharing, with the transition from over-saturation to undersaturation; c). Dependence of the average waiting time (black axis) and average trip time (red axis) on commuter's willingness to taxi-sharing, in the under-saturated region.}
\label{shared}
\end{figure}

Large scale numerical simulations are carried out with a specific route matching algorithm that ensures when ride-sharing occurs, the extra time involved in the detours will not exceed five minutes for any shared parties. We will not go into the details of the simulation, which can be found in Ref\cite{yang}. Both artificial road networks and realistic road networks (with empirical demand for taxi services) are analysed, giving qualitatively similar results. Here we focus on realistic road network and the empirical taxi demand from the city state of Singapore (the road network is shown in Fig.(\ref{singaporemap}). We analyse the dependence of the average waiting time (the time that commuters spend waiting at the nodes) and the average trip time (the time commuters spend inside the taxi) on $p_s$, the percentage of commuters open to ride-sharing. Note that even when a commuter is open to ride-sharing, his/her trip may not be shared with another commuter. This is because at the time such a commuter makes a booking request, there may not be suitable partners to share the trip that satisfies the constraint of the detour time.

As shown in Fig.(\ref{sharephase}), in such systems we also have very well-defined $N^*_{\text{taxi}}$, which decreases with increasing $p_s$. This confirms the intuition that ride-sharing makes the taxi system more efficient. In the oversaturated phase, the average waiting time decreases rapidly with increasing $p_s$; while in the undersaturated phase, the decrease is much more marginal. In both phases, on the other hand, the average trip time increases with $p_s$. This is because with more commuters open to ride-sharing, more detours will be executed, leading to longer trip time. Interestingly, when the number of taxis is large enough, increasing $p_s$ can lead to a phase transition, as one can see from Fig.(\ref{sharephase}) and Fig.(\ref{shared}b). This phase transition leads to a significant drop both in terms of the average waiting time and the average trip time. This is because the phase transition physically corresponds to the transition from the case with no empty taxis roaming, to the case \emph{with} empty taxis roaming. With the latter case, waiting time will be significantly shortened with the booking policy as we have shown in Sec.~\ref{numerics}. With empty taxis roaming, the number of shared trips will also reduce significantly (as in the oversaturated phase with no empty taxis, occupied taxis are the only options for many commuters), leading to fewer detours and lower average trip time.

\section{Discussions and Conclusions}\label{summary}

In conclusion, this paper mainly focuses on theoretical studies of the general dynamics of logistics systems consisting of two types of agents, which we can generically labelled as the ``delivery" agents ($\mathcal L$ agents) and the ``goods" agents ($\mathcal G$ agents). Both agents are generated in the system on a real-time, dynamical basis, and the tasks of the $\mathcal L$ agents are to find $\mathcal G$ agent and deliver them to their respective destinations. We use the taxi system as the main example, whereby the $\mathcal L$ agents are empty taxis, and the $\mathcal G$ agents are commuters. The interaction between the two types of agents are formulated in precise mathematical languages, and the domain of the interaction is over a potentially complex network (e.g. the road network for taxis and commuters). We constructed simple models that capture the essential features of such logistics systems, and solved them analytically, revealing a second order phase transition when the system reaches equilibrium. Large scale agent-based modelling is also carried out for realistic settings and with additional details, revealing a number interesting emergent behaviours that can be well explained with our simple models and the analytical results.

One of the main results is the identification of a second order transition from the oversaturated phase (where supply of $\mathcal G$ agents exceeds that of the $\mathcal L$ agents) and the undersaturated phase (vice versa). The phase boundary is both mathematically well-defined and easily observable in more realistic, finite systems via agent-based simulations. The dynamical behaviours of the logistics systems are qualitatively different in these two phases, and we studied in details on how the phase boundary can be affected by various components of the logistics system. Moreover, the phase boundary gives a proper definition of an optimal number of $\mathcal L$ agents in the system, from the system efficiency point of view. In contrast to some similar definitions in the literature, this optimal number $N^*$ depends entirely on the dynamics over the spatiotemporal distribution of the rate of generation of $\mathcal L$ and $\mathcal G$ agents. It can be well-defined without explicit reference to any specific utility functions we construct over the logistics system. Given its generality, we conjecture that sharp phase transitions can be detected with most of the commonly used utility functions across the same $N^*$. We illustrated this with the taxi systems allowing ride-sharing, where we considered average waiting time and travel time for the commuters, and average trip time which can corresponds to the fuel costs for the drivers.

Even with realistic microscopic simulations performed in this work, there are several important features of the logistics systems which we did not go into details. Taking the taxi systems as an example, while the empty taxis ($\mathcal L$ agent) can move from one node to another in the road network, the passengers ($\mathcal G$ agents) do not move: they always stay at the nodes where they are generated. In reality, passenger behaviours are highly non-trivial: they may move from one node to another, or simply disappear from the road network (e.g. the waiting time is too long). This can be partially captured by making $g_i$ time-dependent, though proper passenger behaviours need to be incorporated in agent-based modelling for higher accuracy. The bottleneck here is the appropriate methodology in constructing mathematical models for the passenger behaviours from the empirical data. In general for the taxi systems, however, we expect most passengers waiting at a fixed location for the taxi services. As long as the waiting time is not too long (so that we are not deep in the oversaturated phase), it is also uncommon for passengers to give up waiting. We thus consider the simulations in this work (where the passenger behaviours are ignored) are good approximations of realistic systems in most situations.

Another important aspect of the realistic traffic network is that the adjacency matrix is not stationary: the edge weights can depend on the flow of traffic through the edge, especially during peak hours when the flow is high. In cases where taxis themselves are not dominating the traffic flow (so the flow mainly consists of other types of vehicles), we can make the adjacency matrix time dependent based on the empirical data, and the simulations carried out in this work can be easily generalised. For slowly varying adjacency matrices, we expect possible transition between undersaturated and oversaturated phases, when the traffic flow becomes congested (i.e. the effective road network becomes larger). This can be captured with simulations using different types of time-independent road networks, one for normal traffic flow, and another for the congested flow. If the adjacency matrix fluctuates rapidly as a function of time, the system dynamics could be more complex. The same is true in cases where the taxis dominate the traffic flow, so the number of taxis in the system can directly impact if certain roads are congested or not. These interesting scenario needs detailed simulation analysis, which could be part of future studies.

Going forward, we will give a more detailed analysis on how to apply the theoretical tools and conceptual development in this work for benchmarking various optimisation schemes for this class of logistics systems, especially when considering different utility functions from the perspectives of customers, operators and policy-makers. Complex logistics systems like the ones discussed in the work can have many components to tune, in order for certain performance indicators to improve. The theoretical treatment in this work allows us to understand both intuitively and quantitatively that given a specific situation, tuning some specific components may lead to greater marginal gain, as we have illustrated in the text right before Sec.~\ref{tuningn}. The existence of the oversaturated phase and the undersaturated phase could be prevalent in many systems similar to the ones we are focusing here. We propose that a thorough analysis of the phase boundary dependence on various factors in the system, with a combination of numerical and analytical tools, should be the first step in optimising dynamical spatiotemporal resource allocation in such systems. While the work focuses exclusively on the decentralised policies of the logistics systems, it would also be interesting to see how a more centralised approach could affect the dynamics of such systems qualitatively. In general, a centralised control of agents can lead to more efficient delivery in principle, but the parameter space for optimisation is much larger. A hybrid of centralised and decentralised policies for $\mathcal L$ agents could be a practical approach for certain logistics systems, which is worth exploring in the future works.
\section*{ACKNOWLEDGMENT}

%%%%%%%%%%%%%%%%%%%%%%%%%%%%%%%%%%%%%%%%%%%%%%%%%%%%%%%%%%%%%%%%%%%%%%%%%%%%%%%%

This research was partially supported by Singapore A$^{\star}$STAR SERC ``Complex Systems" Research Programme grant 1224504056.


\begin{thebibliography}{99}
\bibitem{laporte} Berbeglia, G., Cordeau, J-F. and Laporte, G., 2010. Dynamic pickup and delivery problems. European Journal of Operational Research {\bf 202}, 8-15.
\bibitem{bock} Ferrucci, F. and Bock, S., 2014. Real-time control of express pickup and delivery processes in a dynamic environment, Transportation Research Part B, {\bf 63}, 1-14.
\bibitem{yang} Yang, B., Ren, S., Legara, E.F., Li, Z., Ong, Y.X., Lin, L. and Monterola, C., 2018. Phase Transition in Taxi Dynamics and Impact of Ridesharing, arXiv: 1801.00462.
\bibitem{bilge} Atasoy, B., Ikeda, T., Song X. and Ben-Akiva, M.E., 2015. The concept and impact analysis of a flexible mobility on demand system, Transportation Research Part C, {\bf 56}, 373.
\bibitem{zhong} Sungur, I., Ren, Y., Ordonez, F., Dessouky, M. and Zhong, H., 2010. A Model and Algorithm for the Courier Delivery Problem with Uncertainty, Transportation Science, {\bf 44}, 193-205.
\bibitem{laporte2} Solyali, O., Cordeau, J-F. and Laporte, G., 2011. Robust Inventory Routing Under Demand Uncertainty, Transportation Science, {\bf 46}, 327-340.
\bibitem{jaillet} Flajolet, A., Blandin, S. and Jaillet, P., 2017. Robust Adaptive Routing Under Uncertainty, Operations Research, {\bf 66}, 210-229.
\bibitem{delage} Carlsson, J.G. and Delage, E., 2013. Robust Partitioning for Stochastic Multivehicle Routing, Operations Research, {\bf 61}, 727-744.
\bibitem{chowjy} Djavadian, S. and Chow, J.Y., 2017. An agent-based day-to-day adjustment process for modeling ‘Mobility as a Service’with a two-sided flexible transport market. Transportation research part B, {\bf 104}, 36.
\bibitem{yej} Xu, Z., Yin, Y. and Ye, J., 2019. On the supply curve of ride-hailing systems. Transportation Research Part B, in press.
\bibitem{mohsen} Ramezani, M. and Nourinejad, M., 2017. Dynamic modeling and control of taxi services in large-scale urban networks: A macroscopic approach, Transportation Research Part C, article in press.
\bibitem{reisman} Eksioglu, B., Vural, A.V. and Reisman, A., 2009. The vehicle routing problem: A taxonomic review, Computers and Industrial Engineering, {\bf 57}, 1472.
\bibitem{koening} Furuhata, F., Dessouky, M., Ordóez, F., Brunet, M.E., Wang, X., Koening, S., 2013. Ridesharing: the state-of-the-art and future directions, Transportation Research Part B, \textbf{57} 28.
\bibitem{roorda} Nourinejad, M. and Roorda, M.J., 2016. Agent based model for dynamic ridesharing, Transportation Research Part C, {\bf 64}, 117.
\bibitem{reijers} Li, B., Krushinsky, D., Woensel, T.V. and Reijers, H.A., 2016. The Share-a-Ride problem with stochastic travel times and stochastic delivery locations, Transportation Research Part C, {\bf 67}, 95.
\bibitem{dchen} Farhan, J. and Chen, T.D., 2018. Impact of ridesharing on operational efficiency of shared T autonomous electric vehicle fleet, Transportation Research Part C, {\bf 93}, 310.
\bibitem{wong} Wong, R.C.P., Szeto, W.Y. and Wong, S.C., 2014. Bi-level decisions of vacant taxi drivers traveling towards taxi stands in customer-search: Modeling methodology and policy implications, Transport Policy, {\bf 33}, 73.
\bibitem{damas} Moreira-Matias, L., Gama, J., Ferreira, M. and Damas, L., 2012. A predictive model for the passenger demand on a taxi network, International IEEE Conference on Intelligent Transportation Systems (ITSC), DOI: 10.1109/ITSC.2012.6338680.
\bibitem{yanghai2} Wong, R.C.P., Szeto,  W.Y., Wong, S.C. and Yang, H., 2014. Modelling multi-period customer-searching behaviour of taxi drivers, Transportmetrica B, {\bf 2}, 40.
\bibitem{kendall} Bai, R., Li, J., Atkin, J.A.D. and Kendall, G., 2014. A novel approach to independent taxi scheduling problem based on stable matching, Journal of the Operational Research Society, {\bf 65}, 1501.
\bibitem{ramezani} Nourinejad M. and Ramezani, M., 2016. Developing a large-scale taxi dispatching system for urban networks,  IEEE 19th International Conference on Intelligent Transportation Systems (ITSC), 2016, DOI: 10.1109/ITSC.2016.7795592
\bibitem{chen} Ke, J., Zheng, H., Yang, H. and Chen, X., 2017. Short-term forecasting of passenger demand under on-demand ride services: A spatio-temporal deep learning approach, Transportation Research Part C, {\bf 85}, 591.
\bibitem{wong2} Long, J., Szeto, W.Y., Du, J. and Wong, R.C.P., 2017. A dynamic taxi traffic assignment model: A two-level continuum transportation system approach, Transportation Research Part B, {\bf 100}, 222.
\bibitem{yux} Yu, X., Gao, S., Hu, X. and Park, H., 2019. A Markov decision process approach to vacant taxi routing with e-hailing. Transportation Research Part B, {\bf 121}, 114.
\bibitem{zha} Zha, L., Yin, Y. and Du, Y., 2017. Surge pricing and labor supply in the ride-sourcing market. Transportation Research Procedia, {\bf 23}, pp.2-21.
\bibitem{ham} Ham, A.M., 2018. Integrated scheduling of m-truck, m-drone, and m-depot constrained by time-window, drop-pickup, and m-visit using constraint programming, Transportation Research Part C, {\bf 91}, 1-14.
\bibitem{sundar} Xu, H., Chen, Z-L., Rajagopal, S. and Arunapuram, S., 2003. Solving a Practical Pickup and Delivery Problem, Transportation Science, {\bf 37}, 347-364.
\bibitem{changsun} Cardin, M.A., Deng, Y. and Sun, C., 2017. Real options and flexibility analysis in design and management of one-way mobility on-demand systems using decision rules, Transportation Research Part C, {\bf 84}, 265.
\bibitem{archetti} Nagy, G., Wassan, N.A., Speranza, M.G. and Archetti, C., 2013. The Vehicle Routing Problem with Divisible Deliveries and Pickups, Transportation Science, {\bf 49}, 271-294.
\bibitem{speranza}Archetti, C., Mansini, R. and Speranza, M.G., 2005. Complexity and Reducibility of the Skip Delivery Problem, Transportation Science, {\bf 39}, 182-187.
\bibitem{yangunpublished} Yang, B. et.al. ``Intelligent Turn-by-turn Routing of On-demand Services: Optimal Algorithms based on Reinforcement Learning", work in progress.
\bibitem{bailey}Bailey, N.T.J., 1954. On Queuing Processes with Bulk Services,  Journal of the Royal Statistical Society. Series B (Methodological), {\bf 16}, 80-87.
\bibitem{graph} Spielman, D.A., 2011. Spectral Graph theory. Combinatorial Scientific Computing. Chapman and Hall/CRC Press.
\bibitem{yangli} Yang, B. and Li, Q., 2018. Turn-by-turn Intelligent Manoeuvring of Driverless Taxis: A Recursive Value Model Enhanced by Reinforcement Learning, IEEE Intelligent Vehicles Symposium 2018, in press.
\bibitem{gillespie1977exact} Gillespie, D.T., 1977. Exact stochastic simulation of coupled chemical reactions, The journal of physical chemistry, 81(25) 2340--2361.

\end{thebibliography}
\end{document}